\RequirePackage[2020-02-02]{latexrelease}
\documentclass[conference]{IEEEtran}
\usepackage{cite}
\usepackage{amsmath,amssymb,amsfonts}
\usepackage{algorithmic}

\usepackage{textcomp}
\def\BibTeX{{\rm B\kern-.05em{\sc i\kern-.025em b}\kern-.08em
    T\kern-.1667em\lower.7ex\hbox{E}\kern-.125emX}}
\usepackage{booktabs}
\usepackage{hyperref}  	% Creates hyperlinks in cross references
\hypersetup{
    allcolors=black,
    colorlinks=true,
    linkcolor=black,
    urlcolor=blue,
}
\usepackage{soul}
\usepackage{parskip}
\usepackage{microtype}
\usepackage{multirow}
\usepackage{amsthm}
\usepackage{empheq}
\usepackage{pifont}
\usepackage{float}
\usepackage[section]{placeins}
\theoremstyle{definition}
\newtheorem{definition}{Definition}[section]
\usepackage{empheq}
\usepackage{pifont}
\usepackage{textcomp}
\usepackage{caption}
\usepackage{subcaption}
\usepackage{pgfplots}
\usepackage{pgfplotstable}
\pgfplotsset{compat=1.7}
\usepackage{xcolor}
\usepackage{graphicx}
\usepackage{tikz}
\usepackage[printwatermark]{xwatermark}
\newsavebox\mybox
\savebox\mybox{\tikz[color=black,opacity=0.3]\node{PREPRINT};}
\newwatermark*[
  allpages,
  angle=55,
  scale=17,
  xpos=-70,
  ypos=50
]{\usebox\mybox}
\newcommand\dottedcircle{\tikz \draw [line cap=round, line width=0.20ex, dash pattern=on 0pt off 2pt] (0,0) circle [radius=1ex];}
\newcommand{\rootnode}{\scalebox{2}{$\square$}}
\newcommand{\assetnode}{\scalebox{2}{$\bigcirc$}}
\newcommand{\supernode}{\scalebox{2}{$\dottedcircle$}}
\newcommand{\vulnerabilitynode}{\scalebox{2}{$\blacktriangledown$}}
\newcommand{\dependency}{\scalebox{2}{$\longrightarrow$}}
\newcommand{\update}{\scalebox{2}{$\dashrightarrow$}}
\newcommand*\rot{\rotatebox[origin=c]{90}}

\def\BibTeX{{\rm B\kern-.05em{\sc i\kern-.025em b}\kern-.08em
    T\kern-.1667em\lower.7ex\hbox{E}\kern-.125emX}}

\begin{document}

\title{A Novel Model for Vulnerability Analysis through Enhanced Directed Graphs and Quantitative Metrics}

% \author{\uppercase{\'Angel Longueira-Romero}\authorrefmark{1}, \IEEEmembership{Fellow, IEEE},
% \uppercase{Second B. Author\authorrefmark{2}, and Third C. Author,
% Jr}.\authorrefmark{3},
% \IEEEmembership{Member, IEEE}}

\author{\IEEEauthorblockN{1\textsuperscript{st} Ángel Longueira-Romero, 2\textsuperscript{nd} Rosa Iglesias, 3\textsuperscript{rd} Jose Luis Flores}
\IEEEauthorblockA{\textit{Industrial Cybersecurity} \\
\textit{Ikerlan Technology Research Centre (BRTA)}\\
Arrasate/Mondragón, Spain \\
\{alongueira, riglesias, jlflores\}@ikerlan.es}
\and
\IEEEauthorblockN{4\textsuperscript{th} Iñaki Garitano}
\IEEEauthorblockA{\textit{Dept. of Electronics and Computing} \\
\textit{Mondragon Unibertsitatea}\\
Arrasate/Mondragón, Spain \\
igaritano@mondragon.edu}
}

\maketitle
% =================================================================================
% === 0. ABSTRACT =================================================================
% =================================================================================
\begin{abstract}
% General Background
Industrial components are of high importance because they control critical infrastructures that form the lifeline of modern societies. 
% Specific Background
However, the rapid evolution of industrial components, together with the new paradigm of Industry 4.0, and the new connectivity features that will be introduced by the 5G technology, all increase the likelihood of security incidents. 
These incidents are caused by the vulnerabilities present in these devices. In addition, although international standards define tasks to assess vulnerabilities, they do not specify any particular method. 
%As a result, ICs are subjected to the same type of vulnerabilities as any other system. 
% Statement of the problem or knowledge gap
Having a secure design is important, but is also complex, costly, and an extra factor to manage during the lifespan of the device. 
% ``Here we show...'' one -sentence summary of what was done / learned
This paper presents a model to analyze the known vulnerabilities of industrial components over time. The proposed model is based on two main elements: a directed graph representation of the internal structure of the component, and a set of quantitative metrics that are based on international security standards; such as, the Common Vulnerability Scoring System (CVSS). This model is applied throughout the entire lifespan of a device to track vulnerabilities, identify new requirements, root causes, and test cases. The proposed model also helps to prioritize patching activities. 
% Detailed summary of high-level results
To test its potential, the proposed model is applied to the OpenPLC project. The results show that most of the root causes of these vulnerabilities are related to memory buffer operations and are concentrated in the \textit{libssl} library. Consequently, new requirements and test cases were generated from the obtained data.
\end{abstract}

% De más específico a más genérico

% =================================================================================
% === SECTIONS ====================================================================
% =================================================================================
% =================================================================================
% === I. INTRODUCTION =============================================================
% =================================================================================
\section{Introduction}
\label{sec:introduction}
    % ¿EN DÓNDE VAMOS A CENTRAR EL FOCO DEL ARTÍCULO?
    Industrial components are the driving force of almost every industrial field, such as automotive, manufacturing, telecommunications, energy production, transportation, healthcare, and defense~\cite{ES_Security3, lessonsStuxnet_Chen, ES_Security5, CriticalInfrastructures, SpaceEmbeddedDEvice, intro_5}. These types of components are rapidly evolving~\cite{IIoT_Conectivity, review_lowMiddleHigh_end} and are rapidly increasing in number~\cite{intro_4}:

    \begin{itemize}
        \item Open-source hardware and software, and Commercial Off-The-Shelf (COTS) components are being integrated to speed up their development.
        \item They are increasingly connected, providing more advanced connectivity features, enabling new automation applications, services, and data exchange.
        \item The complexity of industrial systems is also increasing, thus the complexity of their software and hardware is also increasing.
    \end{itemize}

    % OFF-THE-SELF COMPONENTS + OPEN SOURCE
    The reuse of open-source software components is a \textit{de facto} industry norm, with 90\% of the participants using pre-existing code~\cite{imanol_1, imanol_2, imanol_3, sonatype}. Moreover, Commercial Off-The-Shelf (COTS) components are highly available in the market, which makes them suitable for speeding up the development of industrial components~\cite{reviewer1_3}. COTS components are used for hardware, software, and communication interfaces. Thus, vulnerabilities within such components can create potential entry points for malicious adversaries aiming to disrupt CPS operations~\cite{cots_1}. In addition, the use of COTS components makes it easier than ever for industrial components to connect among them and to the Internet, increasing their attack surface even further~\cite{cots_2}.
    
    % Idea de han pasado de NO estar conectados a estar conectados
    % CONECTIVIDAD = INDUSTRIA 4.0 + IoT + IIoT + AI + 5G
    Interconnected systems significantly increase the exposure to many security risks, with critical, environmental and wellbeing impacts. The fifth generation (5G) of wireless technology for cellular networks will facilitate an enhanced Internet connectivity and accommodate the connection of multiple devices through the IoT architecture, which will open further the window of exposure to any threat~\cite{5G1, intro_4, intro_5, intro_6}. Furthermore, this increased connectivity is part of the new paradigm of Industry 4.0. Thus, technologies such as the Internet of Things (IoT)~\cite{intro_1, intro_2, intro_3, intro_6}, cloud computing, Artificial Intelligence (AI)~\cite{intro_3, xin_2018}, and big data (among others) are being extensively used~\cite{industria2, industria3, industria4, industria5, industria6, industria7}, which will again open even further the door for potential attacks.
    
    % COMPLEXITY
    Industrial components are working on an evolving ecosystem that becomes more complex over time: the use of COTS components and open-source software, the enhanced connectivity among them and to the Internet, and the integration of technologies such as IoT, AI, and big data. To face this scenario, industrial components are also becoming more complex. Complexity is a critical aspect of industrial components design, because it is closely related to the number of vulnerabilities~\cite{complexity_1, complexity_2}.

    % SUMMARY
    Security is turning into a key issue in an environment where security is traditionally addressed as Security through obscurity, and treated as an add-on feature instead of being a priority during the development stage~\cite{MatheuSara, NIST_800_82, enisa_SCADA, whatYouCrash_ES}. Numerous attacks have been reported targeting industrial enterprises across the globe since 2010~\cite{dissectingStuxnet_Langner}, and an exponential rise in such attacks is predicted for the upcoming years~\cite{graphNased_Security_Framework_2017, attacksEmbedded_Papp}.

    % DESCRIBIR LA MOTIVACIÓN DE NUESTRO TRABAJO (GAPS)
    % POR QUÉ ES NECESARIO
    % Prioridad
    % Coste si se conoce la gravedad
    % Definición de una estrategia
    % Cumplimiento de la normativa
    
    Under this scenario, a model for continuous vulnerability assessment is needed to manage the security of existing industrial components during their lifespan to deal with emerging threats~\cite{isecomOSSTMM_PDF, owaspGuide}. That is to say, performing a vulnerability analysis at a single point in time (\textit{e.g.}, during development or when a product has been released) is not enough for industrial components and their large lifespan has to be considered~\cite{KLEIDERMACHER20121, kasperskyLab}. Such a model should be able to compute the number and distribution of vulnerabilities among the assets of an industrial component, to detect and classify security vulnerabilities for further remediation or mitigation actions. Moreover, this model should be able to integrate other metrics, such as the severity value of vulnerabilities. 
    % ¿Eliminar esto?
    Severity is closely related to risk assessment and threat model activities. So this value can help in early steps (\textit{e.g.}, design). By incorporating information on their severity, more metrics could be developed to track and enhance the development of industrial components, which will create a more precise threat model~\cite{Thomas_CategorisingWeaknesses_2020}.
    % ¿Eliminar hasta aquí?
    Metrics help in prioritizing the patching of vulnerabilities, obtaining their root causes, and monitoring the evolution of industrial components, not only during their development, but also throughout their lifespan. Finally, the model should be aligned with the most relevant cybersecurity standards to enhance evaluation tasks that they propose or even to cover gaps. Furthermore, both software and hardware should be evaluated because it is of high importance in industrial components, given that the strong bonding between hardware and software is an intrinsic feature of industrial components~\cite{ES_Security, ES_Security2, Marwedel_2018, reviewer1_3}.

    Although great efforts are being made to develop new and better ways to analyze vulnerabilities~\cite{imanol_4, imanol_5}, to measure them (\textit{e.g.}, Common Vulnerabilities and Exposures (CVE), Common Vulnerability Scoring System (CVSS)~\cite{CVE_Definitions, CVE1, CVE2, CVSS1}, or Common Weakness Enumeration (CWE)~\cite{CWE0, CWE1}, or to aggregate them \cite{dbAggregation}), to the best of our knowledge, existing models do not cover industrial components. To set the first steps toward filling this gap, this research work proposes a model with the aim of performing a continuous vulnerability assessment to determine the source and nature of vulnerabilities, and enhance the security of industrial components.

    % De lo más general a lo más específico: metodología = modelo + métricas --> Luego, explicar piezas
    The proposed model is built on top of a directed graph-based structure, and a set of metrics based on globally accepted security standards. This model is intended to be used to help manage the analysis of known vulnerabilities, their root causes, and their impact on the entire life cycle of an industrial component. The internal structure, in terms of assets, of an industrial component is represented using directed graphs: different types of nodes represent vulnerabilities for both software and hardware assets. For further analysis, metrics were developed integrating internationally recognized security standards, such as Common Platform Enumeration (CPE)~\cite{CPE_main, CPE7695, CPE7696}, CVE~\cite{CVE1}, CVSS~\cite{CVSS1}, CWE~\cite{CWE0}, and Common Attack Pattern Enumeration and Classification (CAPEC)~\cite{CAPEC1, CAPEC_Definitions}. By using the proposed metrics, the evaluator is capable of identifying the source and nature of the detected vulnerabilities, and help in prioritizing their patching. Finally, the work presented here is also aligned with the ISA/IEC 62443 standard ``Security for industrial automation and control systems''.

    This paper is structured as follows: 
        First, the related work is reviewed in Section \ref{sec:relatedWork}. Then, the main pieces of the proposed model are defined in Section \ref{sec:introModel}. Second, to demonstrate the potential of this proposal, the proposed model is applied to a real use case (Section \ref{sec:usecase}. Finally, conclusions and future work of this research are described in Section \ref{sec:conclusionsFutureWork}.
% \input{docs/2.Background}
% =================================================================================
% === II. Related Work ============================================================
% =================================================================================
\section{Related Work}
\label{sec:relatedWork}
    This section will review the current status of vulnerability assessment. This review aims to find similar approaches from the literature, including the current standard and metrics.

% Industry standards are a set of criteria within an industry that establish requirements for products, practices, or operations in a given field. Standards are commonly used by industry,  and it is important to take them into account,

    \subsection{Vulnerability Analysis in Security Standards}
        Industry is currently making a significant effort to incorporate security aspects into the development of industrial components, which has led to a set of standards, such as the ISA/IEC 62443. ISA/IEC 62443 is a family of standards which includes several parts, where Part 4 is focused on the specific development of components (\textit{e.g.}, embedded systems, host systems, network devices, and software applications). This standard, in which certain parts are not yet fully defined, has been inspired by previous standards that have a long tradition of use, such as the Common Criteria. Consequently, this section will not only describe the ISA/IEC 62443 standard, but will also analyze the Common Criteria. This review is focused on how these standards conduct vulnerability analysis, the use of metrics, their management of the life cycle of the device, the techniques that they propose, and the security evaluation of both software and hardware.

        \subsubsection{ISA/IEC 62443}
            Based on the ISA-99 document, the ISA/IEC 62443 constitutes a series of standards, technical reports, and related information that define the procedures and requirements for implementing electronically secure Industrial Automation and Control Systems (IACSs)~\cite{IEC62443}. As expressed by this standard, security risk management shall jointly and collaboratively be addressed by all the entities involved in the design, development, integration, and maintenance of the industrial and/or automation solution (including subsystems and components) to achieve the required security level~\cite{securityIssues_Mugarza_62443_2020}.

            % Párrafo de conexión entre estos dos
            This joint effort is reflected in the organization of the documents of the standard, which is divided into four parts:
            \begin{enumerate}
                \item Part 1 - General: Provides background information such as security concepts, terminology and metrics;
                \item Part 2 - Policies and procedures: Addresses the security and patch management policies and procedures;
                \item Part 3 - System: Provides system development requirements and guidance;
                \item Part 4 - Component: Provides product development and technical requirements, which are intended for product vendors.
            \end{enumerate}

            The ISA/IEC 62443-4-1 technical document is divided into eight practices, which specify the secure product development life cycle requirements for both the development and the maintenance phases~\cite{IEC62443_4_1}.
            % ¿Hace análisis de vulnerabilidades?
            ``Practice 5 - Security verification and validation testing'' (SVV) section of this document specifies that a process shall be employed to identify and characterize potential security vulnerabilities in the product, including known and unknown vulnerabilities~\cite{BasicConcepts_2004, zeroDayVulnerabilities_2019}. Two requirements in Practice 5 are in charge of the task of analyzing vulnerabilities, as follows:
    
            % ¿Cómo hace el análisis de vulnerabilidades?
            \begin{itemize}
                \item{Requirement SVV-3. Vulnerability Testing~\cite{IEC62443_4_1}}. This requirement states that a process shall be employed to perform tests that focus on identifying and characterizing potential and known security vulnerabilities in the product (\textit{i.e.}, fuzz testing, attack surface analysis, black box known vulnerability scanning, software composition analysis, and dynamic runtime resource management testing).
            
                \item{Requirement SVV-4. Penetration Testing~\cite{IEC62443_4_1}}. This requirement states that a process shall be employed to identify and characterize security-related issues via tests that focus on discovering and exploiting security vulnerabilities in the product (\textit{i.e.}, penetration testing).
            \end{itemize}
    
            % ¿Tiene en cuenta las dependencias entre vulnerabilidades y activos? ¿Cómo? ¿Usa árboles de dependencias u otro tipo de grafos para ello?
            Although the ISA/IEC 62443-4-1 document considers the possibility of analyzing and characterizing the vulnerabilities of an industrial component, it does not propose a technique to perform this task, but instead refers to other standards for vulnerability handling processes~\cite{ISO30111}. In addition, it does not indicate how the data obtained from the analysis should be interpreted, and it does not define metrics or reference values for the current state of compliance with the requirement. Finally, it does not take into account neither the dependencies among the assets of the industrial component (dependency trees), nor their evolution of the number of vulnerabilities over time.

            % ¿Propone métricas y fomenta su uso?
            % ¿La evaluación contempla tanto el SW como el HW? That said, neither the CC nor the ISA/IEC 62443 evaluates both dimensions of the IC, and more emphasis is placed on software. 
            % ¿Cómo se gestionan las actualizaciones?
            % ¿Contempla todo el ciclo de vida del producto?

        \subsubsection{Common Criteria}
            The Common Criteria (CC) for Information Technology Security Evaluation (ISO/IEC 15408) is an international standard that has a long tradition in computer security certification~\cite{commonCriteriaGeneralModel}. CC is a framework which provides assurance that the processes of specification, implementation, and evaluation of a computer security product have been conducted in a rigorous, standard, and repeatable manner at a level that is commensurate with the target environment for use.

            To describe the rigor and depth of an evaluation, the CC defines seven Evaluation Assurance Levels (EALs) on an increasing scale~\cite{commonCriteriaGeneralModel}, from EAL1 (the most basic) to EAL7 (the most stringent security level). It is important to notice that the EAL levels do not measure security itself. Instead, emphasis is given to functional testing, confirming the overall security architecture and design, and performing some testing techniques (depending on the EAL to be achieved).
    
            % ¿Cómo hace el análisis de vulnerabilidades?
            The CC defines five tasks in the Vulnerability Assessment class, which manage the deepness of the vulnerability assessment. The higher the EAL to be achieved, the greater the number of tasks in the list to be performed~\cite{commonCriteria_Part3}:

                \begin{enumerate}
                    \item Vulnerability survey,
                    \item Vulnerability analysis,
                    \item Focused vulnerability analysis,
                    \item Methodical vulnerability analysis, and 
                    \item Advanced methodical vulnerability analysis.
                \end{enumerate}

            Every task checks for the presence of publicly known vulnerabilities. Penetration testing is also performed. The main difference among the five levels of vulnerability analysis described here is the deepness of the analysis of known vulnerabilities and the penetration testing.

            % ¿Tiene en cuenta las dependencias entre vulnerabilidades y activos? ¿Cómo? ¿Usa árboles de dependencias u otro tipo de grafos para ello?
            The CC scheme defines the general activities, but it does not specify how to perform them, therefore no technique for analyzing vulnerabilities is proposed. The evaluator decides the most appropriated techniques for each test in each scenario  and for each device, which adds a large degree of subjectivity to the evaluation. Furthermore, dependencies among vulnerabilities and assets are not considered in the analysis. Moreover, the CC does not define a procedure to manage the life cycle of the device. In other words, when updated, the whole device has to be reevaluated~\cite{usingCC, MatheuSara, melladoCC, CCupdate}. Finally, although the usage of metrics is encouraged by the CC, it does not propose any explicitly defined metric to be used during the evaluation. 
            % Finally, the evaluation is a costly process, and the effort and time necessary to prepare evaluation evidence and other evaluation-related documentation is so cumbersome that by the time the work is completed, the product under evaluation is generally obsolete~\cite{MatheuSara, CCAnalysis}.

            % ¿Propone métricas y fomenta su uso?
            % ¿La evaluación contempla tanto el SW como el HW? That said, neither the CC nor the ISA/IEC 62443 evaluates both dimensions of the IC, and more emphasis is placed on software. 
            % ¿Cómo se gestionan las actualizaciones?
            % ¿Contempla todo el ciclo de vida del producto?

    \subsection{Vulnerability Analysis Methodologies}
        Vulnerability analysis is a key step towards the security evaluation of a device. Consequently, many research efforts have been focused on solving this issue. In this subsection, the most relevant works related to vulnerability analysis are reviewed.
    
        Homer \textit{et al.}~\cite{homer2009sound} present a quantitative model for computer networks that objectively measures the likelihood of a vulnerability. Attack graphs and individual vulnerability metrics, such as CVSS, and probabilistic reasoning are applied to produce a sound risk measurement. However, the main drawback is that their work is only applicable to computer networks. 
        Although they propose new metrics based on the CVSS for probabilistic calculations, they do not integrate standards such as CAPEC to enhance their approach centered on possible attacks and privilege escalation. They also fail to establish a relationship among existing vulnerabilities, and they fail to obtain the source problem causing each vulnerability.

        Zhang \textit{et al.}~\cite{zhang2011empirical, homer2013aggregating} developed a quantitative model that can be used to aggregate vulnerability metrics in an enterprise network based on attack graphs. Their model measures the likelihood that breaches can occur within a given network configuration, taking into consideration the effects of all possible interplays between vulnerabilities. 
        This research is centered on computer networks, using attack graphs. Although the proposed model is capable of managing shared dependencies and cycles, only CVSS-related metrics are used. Moreover, this model assumes that the attacker knows all of the information in the generated attack graphs. Finally, the method that they proposed for the aggregation of metrics is only valid for attack graphs, and is not valid for vulnerability analysis. 
        % This model also fails to integrate the configuration of the system in the analysis.
        
        George \textit{et al.}~\cite{graphBasedSecurityIoT_2018} propose a graph-based model to address the security issues in Industrial IoT (IIoT) networks. Their model is useful because it represents the relationships among entities and their vulnerabilities, serving as a security framework for the risk assessment of the network. Risk mitigation strategies are also proposed. Finally, the authors discuss a method to identify the strongly connected vulnerabilities. However, the main drawback of this work is that each node of the generated attack graph represents a vulnerability instead of representing a device or an asset of that device. This leads to a loss of information in the analysis, because there is no way to know which vulnerability belongs to which device. Moreover, these methods need to know the relationships among present vulnerabilities in the devices. This information is not trivially obtained, and a human in the loop is needed. The proposals of~\cite{reviewer1_1} and~\cite{reviewer1_2} follow a similar graph-based approach to study the effects of cascade failures in the power grid, and a subway network.
        
        Poolsappasit \textit{et al.}~\cite{dynamicRiskManagement_2012} propose a risk management framework using Bayesian networks that enables a system administrator to quantify the chances of network compromise at various levels. The authors are able to model attacks on the network, and also to integrate standardized information of the vulnerabilities involved, such as their CVSS score. Although their proposed model lends itself to dynamic analysis during the deployed phase of the network, these results can only be applied to computer networks. Meanwhile, the prior probabilities that are used in the model are assigned by network administrators, and hence are subjective. The proposed model also has some issues related to scalability. 
        % The system configuration is also not considered.

        Mu\~noz-González \textit{et al.}~\cite{exactInferenceTechniques_2019} propose the use of efficient algorithms to make an exact inference in Bayesian attack graphs, which enables static and dynamic network risk assessments. This model is able to compute the likelihood of a vulnerability, and can be extended to include zero-day vulnerabilities, attacker’s capabilities, or dependencies between vulnerability types. 
        Although this model is centered on studying possible attacks, it fails to integrate standards (such as CAPEC) that are related to attack patterns. Moreover, the generated graphs are focused on privilege escalation, trust, and users, rather than including information about vulnerabilities and the analyzed device.
        
        Liu \textit{et al.}~\cite{Liu_2019} carry out a detailed assessment of vulnerabilities in IoT-based critical infrastructures from the perspectives of applications, networking, operating systems, software, firmware, and hardware. They highlight the three key critical infrastructure IoT-based cyber-physical systems (\textit{i.e.}, smart transportation, smart manufacturing, and smart grid). They also provide a broad collection of attack examples upon each of the key applications. Finally, the authors provide a set of best practices and address the necessary steps to enact countermeasures for any generic IoT-based critical infrastructure system. Nevertheless, their proposal is focused on attacks and countermeasures, and it leaves aside the inner analysis of the targets. Continuous evaluation over time is not considered in this proposal, and no enhancements of the development process are generated.

        Hu \textit{et al.}~\cite{Hu_2020} Hu et al. propose a network security risk assessment method that is based on the improved hidden Markov model (I-HMM). The proposed model reflects the security risk status in a timely and intuitive manner, and it detects the degree of risk that different hosts pose to the network. Although this is a promising approach, it is centered on computer networks and is at a higher abstraction level. No countermeasure or enhancement in the development process is proposed or generated.

        Zografopoulos \textit{et al.}~\cite{Zografopoulos_2021} provide a comprehensive overview of the Cyber-Physical System (CPS) security landscape, with an emphasis on Cyber-Physical Energy Systems (CPES). Specifically, they demonstrate a threat modeling methodology to accurately represent the CPS elements, their interdependencies, as well as the possible attack entry points and system vulnerabilities. They present a CPS framework that is designed to delineate the hardware, software, and modeling resources that are required to simulate the CPS. They also construct high-fidelity models that can be used to evaluate the system's performance under adverse scenarios. The performance of the system is assessed using scenario-specific metrics. Meanwhile, risk assessment enables system vulnerability prioritization, while factoring the impact on the system's operation. Although this research work is comprehensive, it is focused on enhancing the existing adversary and attack modeling techniques of CPSs of the energy industry. Moreover, their model does not integrate the internal structure of the target of evaluation, and it does not take both software and hardware into account for the evaluation. Continuous evaluation over time is not considered. Finally, they do not propose countermeasures, or any kind of mechanism to enhance the security or the development of the CPSs.
        
        Most of the works reviewed here are more focused on modeling threats and attacks. They do not propose solutions to protect CPSs, enhancing their development, or manage their update throughout their whole life cycle. It is worth noting that they are still more focused on software evaluation, while hardware is usually neglected in their proposals.
        
        As shown in this review, most of the research has adopted dependency trees, attack graphs, or directed graphs as the main tool to manage and assess vulnerabilities in computer networks. Graphs are an efficient technique to represent the relationships between entities, and they can also effectively encode the vulnerability relations in the network. Furthermore, the analysis of the graph can reveal the security-relevant properties of the network. For fixed infrastructure networks, graphical representations, such as attack graphs, are developed to represent the possible attack paths by exploiting the vulnerability relationships. For these reasons, vulnerability analysis techniques based on directed graphs are frequently found in the literature~\cite{bayesianDesicion_2020}. However, despite their potential, these analysis techniques have been relegated to vulnerability analysis in computer networks. Graph-based analysis has rarely been applied to industrial components.

    \subsection{Security Metrics}
        Standards of measurement and metrics 
        %Metrics\footnote{A measurement is a concrete and objective attribute that provides a single-point-in-time view of a specific and discrete factor; whereas, a metric is generated from the analysis of the raw data provided by measurements.} 
        are a powerful tool to manage security and for making decisions~\cite{whySecurityMetrics_Atzeni, Zeb_QuantitativeSecurityMetrics_2018, securityMetrics_indin2020}. If carefully designed and chosen, metrics can provide a quantitative, repeatable, and reproducible value. This value is selected to be related to the property of interest of the systems under test (\textit{e.g.}, number and distribution of vulnerabilities). The use of metrics enables results to be compared over time, and among different devices. In addition, they can also be used to systematically improve the security level of a system, or to predict this security level in a future point in time.
        
        Although the capabilities of metrics have been demonstrated, they are not free of drawbacks. In our previous research work~\cite{securityMetrics_indin2020}, we performed a systematic review of the literature and standards. To detect possible gaps, our objective was to find which types of metrics have been proposed and in which fields have been applied.
        % Métricas en estándares
        This research work concludes that, in general, standards encourage the use of metrics, but they do not usually propose any specific set of metrics. If metrics are proposed, then they are conceived to be applied at a higher level (\textit{i.e.}, organization level), and then cannot be applied to industrial components. This type of metric is usually related to measuring the return on security investment, security budget allocation, and reviewing security related documentation.
        
        % Métricas en la literatura
        Our previous results also highlight that scientific papers have focused their efforts on software-related metrics: $77.5 \%$ of the analyzed metrics were exclusively applicable to software (\textit{e.g}., lines of code, number of functions and so on), whereas only $0.6 \%$ were related exclusively to hardware (\textit{e.g.}, side-channel vulnerability factor metric). In addition, $14.8 \%$ of them could be applied to both software and hardware (\textit{e.g.}, the historically exploited vulnerability metric that measures the number of vulnerabilities exploited in the past), and the remaining $7.1 \%$ are focused on other aspects, such as user usability. This shows that there is a clear lack of hardware security metrics in the literature, and the main contributions are centered in software security.
        
        % Gaps detectados
        Other research works also reveal common problems across security metrics~\cite{criticalSecurityIndicators_Rudolph, Sentilles2018WhatDW}:
        \begin{itemize}
            \item Hardly any security metric has a solid theoretical foundation or empirical evidence in support of the claimed correlation.
            \item Many security metrics lack an adequate description of the scale, unit, and reference values to compare and interpret the results. 
            \item Only a few implementations or programs were available to test these security metrics, and only one of the analyzed papers performed some kind of benchmarking or comparison with similar metrics.
            \item The information provided in the analyzed papers is insufficient to understand whether the proposed metrics are applicable in a given context, or how to use them.
        \end{itemize}
                
        Under this scenario, it seems reasonable that future research should be focused on the development of a convincing theoretical foundation, empirical evaluation, and systematic improvement of existing approaches, in an attempt to solve the lack of widely-accepted solutions. In this research work, metrics constitute a key element. They are developed to analyze the distribution of vulnerabilities and to track their evolution over time.
% =================================================================================
% === IV. Description of the Proposed Model =======================================
% =================================================================================
\section{Proposed Approach}
\label{sec:introModel}
    In this research work, we propose a model for the continuous assessment of vulnerabilities over time in industrial components. The proposed model is intended to:

    \begin{itemize}
        \item Identify the root causes and nature of vulnerabilities, which will enable the extraction of new requirements and test cases.
        % \item Extract new requirements and test cases.
        \item Support the prioritization of patching.
        \item Track vulnerabilities during the whole lifespan of industrial components.
        \item Support the development and maintenance of industrial components.
    \end{itemize}

    To accomplish this task, the proposed model comprises two basic elements: the model itself, which is capable of representing the internal structure of the system under test; and a set of metrics, which allow conclusions to be drawn about the origin, distribution, and severity of vulnerabilities. Both the model and metrics are very flexible and exhibit some properties that make them suitable for industrial components, and can also be applied to enhance the ISA/IEC 62443 standard.

    The content in this section is distributed into four sections, namely:
    \begin{enumerate}
        \item \textbf{Model:} The proposed model is explained, together with the systems in which it can be applied and the algorithms that are used to built it. 
        \item \textbf{Metrics:} Metrics are a great tool to measure the state of the system and to track its evolution. The proposed metrics and their usage are described in this section.
        \item \textbf{Properties:} The main features of the proposed model and metrics (\textit{e.g.}, granularity of the analysis, analysis over time, and patching policy prioritization support) are described in detail.
        \item \textbf{Applicability:} Even though the reviewed standards exhibit some gaps, the proposed model aims to serve as the first step towards generating a set of tools to perform a vulnerability analysis in a reliable and continuous way. This last section will discuss the requirements of the ISA/IEC 62443-4-1 that can be enhanced using our model.
    \end{enumerate}

    % \begin{enumerate}
    %     \item \textbf{Description of the model:} The model and its basic elements are defined. The systems on which it can be applied are also described.
    %     \item \textbf{Construction of the model:} The algorithms and steps to build the model are explained.
    %     \item \textbf{Security metrics:} Metrics are a great tool to measure the state of the system and track its evolution. The proposed metrics and their usage is described in this section.
    %     \item \textbf{Properties:} The main features of the proposed model and metrics (\textit{e.g.}, granularity of the analysis, analysis over time, and patching policy prioritization support) are described in detail.
    %     \item \textbf{Applicability:} Even though reviewed standards exhibit some gaps, the proposed model aims to serve as the first step towards generating tools to perform a vulnerability analysis in a reliable and continuous way. In this last section, the requirements of the ISA/IEC 62443-4-1 that can be enhanced using our model are discussed.
    % \end{enumerate}
% =================================================================================
% === V. DESCRIPTION OF THE PROPOSED MODEL ========================================
% =================================================================================
% \section{A Novel Methodology For Vulnerability Analysis}
    \subsection{Description of the Model}
    \label{sec:modelDescription}
        The model that is proposed in this research work is based on directed graphs, and requires knowledge of the internal structure of the device to be evaluated (i.e., the assets, both hardware and software, that comprise it and the relationships between them). This section defines the most basic elements that make up the model, the algorithms to build it for any give system, and its graphical representation.

% ----------------------------------------
% --------- EXTENDED DEPENDENCY GRAPH ----
% ----------------------------------------

        \begin{definition}
            A System Under Test (SUT\footnote{Following the denomination in the ISA/IEC 62443 standard~\cite{IEC62443}, the SUT may be an industrial component, a part of an industrial component, a set of industrial components, a unique technology that may never be made into a product, or a combination of these.}) is now represented by an Extended Dependency Graph (EDG) model $G = \left( \langle A, V \rangle, E \right)$ that is based on directed graphs, where $A$ and $V$ represent the nodes of the graphs, and $E$ represents its edges or dependencies:

            \begin{itemize}
                \item $A = \{a_1, ..., a_n\}$ represents the set of assets in which the SUT can be decomposed, where $n$ is the total number of obtained assets. An asset $a$ is any component of the SUT that supports information-related activities, and includes both hardware and software~\cite{MAGERIT, enisa_glossary, ISO_13335_2004}. Each asset is characterized by its corresponding CPE identifier, while its weaknesses are characterized by the corresponding CWE identifier. In the EDG model, the assets are represented by three types of nodes in the directed graphs (\textit{i.e.}, root nodes, asset nodes and cluster).

                \item $V = \{v_1, ..., v_q\}$ represents the set of known vulnerabilities that are present in each asset of $A$, where $q$ is the total number of vulnerabilities. They are characterized by the corresponding CVE and CVSS values. 
                In the EDG model, vulnerabilities are represented using two types of nodes in the directed graphs (\textit{i.e.}, known vulnerability nodes and clusters).

                \item $E = \{e_{ij} | \forall i, j \in \{1, ..., n+q\} \text{ such that } i \neq j \}$ represents the set of edges or dependencies among the assets, and between assets and vulnerabilities. $e_{ij}$ indicates that a dependency relation is established from asset $a_i$ to asset $a_j$. Dependencies are represented using two different types of edges in the EDG (\textit{i.e.}, normal dependency and deprecated asset/updated vulnerability edges).
            \end{itemize}
        \end{definition}

        In other words, the EDG model can represent a system, from its assets to its vulnerabilities, and its dependencies as a directed graph. Assets and vulnerabilities are represented as nodes, whose dependencies are represented as arcs in the graph. The information in the EDG is further enhanced by introducing metrics.
        
        The EDG model of a given SUT will include four types of node and two types of dependency. The graphical representation for each element is shown in Table \ref{tab:Syntax}. Fig. \ref{fig:EDG} shows an example of a simple EDG and its basic elements. All of the elements that make up an EDG will be explained in more detail below:
        
        % =======
        % FIG. XX
        % =======
        \begin{figure}[!htb]
          \begin{center}
          \includegraphics[width=3.4in]{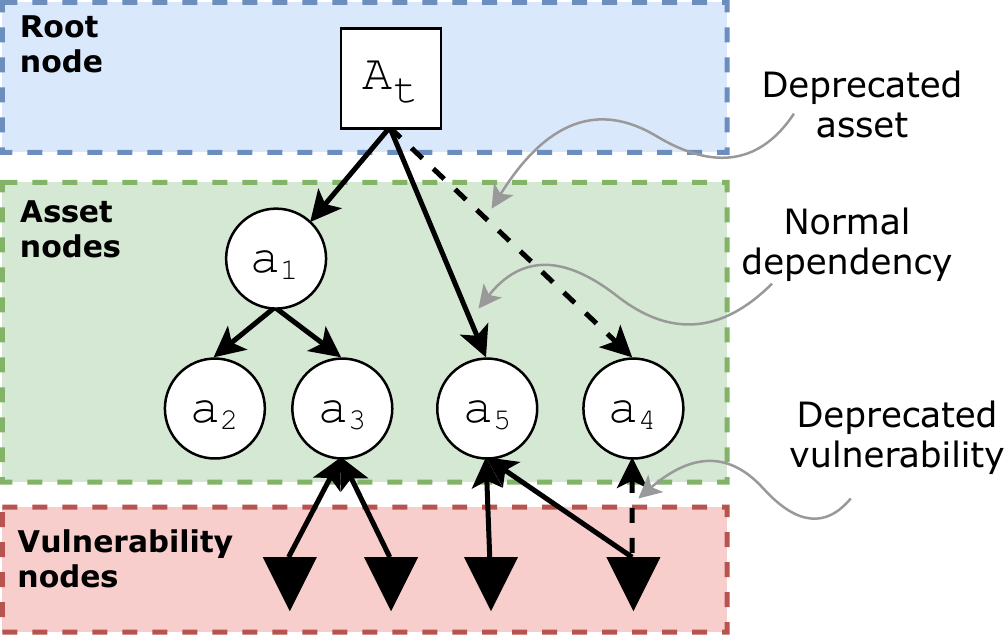}
          \caption{Basic elements of an EDG. Note that clusters are not displayed in this figure. For clusters, see Fig. \ref{fig:supernode_CriteriaapplicationExamples}. For metrics definition, see Section \ref{sec:securityMetrics}.}
          \label{fig:EDG}
          \end{center}
        \end{figure}
        
        \begin{table*}[!htb]
        \centering
        \caption{Overview of the information that is necessary to define each of the EDG elements.}
        \label{tab:Syntax}
        \resizebox{6.8in}{!}{%
        \begin{tabular}{@{}ccll@{}}
        \toprule
        \multicolumn{1}{l}{SYMBOL} & \multicolumn{1}{l}{NOTATION} & MEANING & VALUES                                          \\ \midrule
        $\rootnode$                & $A(t)$   & \begin{tabular}[c]{@{}l@{}}Root Node / \\ Device Node\end{tabular}              & $ CPE_{current} $                               \\
                                   &          &                                                                                 &                                                 \\
        $\assetnode$               & $a(t)$   & Asset Node                                                                      & $ CPE_{previous}, CPE_{current}, CWE_{a_i}(t) $ \\
                                   &          &                                                                                 &                                                 \\
        $\supernode$ & $\underline{a}(t)$ & cluster & $ \{ CPE_{previous}, CPE_{current}, CWE_{a_i}(t) \}, \{ CVE_{a_i}(t), CVSS_{v_i}(t), CAPEC_{w_i}(t) \}, \{Dependencies\} $ \\
                                   &          &                                                                                 &                                                 \\
        $\vulnerabilitynode$       & $v(t)$   & Known Vulnerability Node                                                        & $ CVE_{a_i}(t), CVSS_{v_i}(t), CAPEC_{w_i}(t) $ \\
                                   &          &                                                                                 &                                                 \\
        $\dependency$              & $e(t)$   & Dependency Relation                                                             & ---                                             \\
                                   &          &                                                                                 &                                                 \\
        $\update$                  & $e(t)$   & \begin{tabular}[c]{@{}l@{}}Updated Asset /\\ Patched Vulnerability\end{tabular} & ---                                             \\ \bottomrule
        \end{tabular}%
        }
        \end{table*}

        \subsubsection{Types of Node}
            The EDG model uses four types of node: 
            \begin{itemize}
                \item \textbf{Root nodes} represent the SUT, 
                \item \textbf{Asset nodes} represent each one of the assets of the SUT, 
                \item \textbf{Known vulnerability nodes} represent the vulnerabilities in the SUT, and
                \item \textbf{Clusters} summarize the information in a subgraph.
            \end{itemize}

            \textbf{Root nodes} (collectively, set $G_{R}$) are a special type of node that represent the whole SUT. Any EDG starts in a root node and each EDG will only have one single root node, with an associated timestamp $(t)$ that indicates when the last check for changes was done. This timestamp is formatted following the structure defined in the ISO 8601 standard for date and time~\cite{ISO_2019_IDE}.

            \textbf{Asset nodes} (collectively, set $G_{A}$) represent the assets that comprise the SUT. The EDG model does not impose any restrictions on the minimum number of assets that the graph must have. However, the SUT can be better monitored over time when there is a higher number of assets. Moreover, the results and conclusions obtained will be much more accurate. Nevertheless, each EDG will have as many asset nodes as necessary, and the decomposition of assets can go as far and to as low-level as needed.
            
            Each known vulnerability node will be characterized by the following set of values:

            \begin{itemize}
                \item $CPE_{current}$: Current value for the CPE. This points to the current version of the asset it refers to.
                \item $CPE_{previous}$: Value of the CPE that identifies the previous version of this asset. This will be used by the model to trace back all the versions of the same asset over time, from the current version, to the very first version. 
                \item $CWE_{a_i}(t)$: Set of all the weaknesses that are related to the vulnerabilities present in the asset. The content of this list can vary depending on the version of the asset.
            \end{itemize}
            
\begin{figure}[!htb]
              \begin{center}
              \includegraphics[width=3.4in]{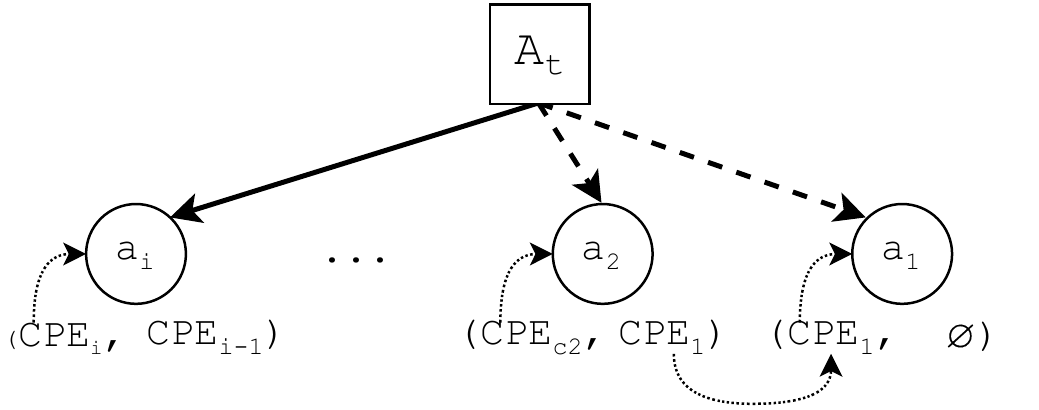}
              \caption{Tracking dependencies between the previous and current CPE values for asset $a$.}
              \label{fig:cpe_relations}
              \end{center}
            \end{figure}

            Fig. \ref{fig:cpe_relations} illustrates how the tracking of the versions of an asset using CPE works. On the one hand, version $a_{i}$ is the current version of asset $a$. It contains its current CPE value, and the CPE of its previous version. On the other hand, $a_{2}$ and $a_{1}$ are previous versions of asset $a$. The last value of $a_1$ points to a null value. This indicates that it is the last value in the chain, and therefore the very first version of the asset $a$.
    
            % =======
            % FIG. XX
            % =======

            \textbf{Known vulnerability nodes} (collectively, set $G_{V}$) represent a known vulnerability present in the asset that it relates to. Each asset will have a known vulnerability node for each known vulnerability belonging to that asset. Assets alone cannot tell how severe or dangerous the vulnerabilities might be, so unique characterization of vulnerabilities is crucial~\cite{graphBasedSecurityIoT_2018}.
    
            To identify each known vulnerability node, each will be characterized by the following set of features (formally defined in \ref{sec:securityMetrics}:
    
            \begin{itemize}
                \item $CVE_{a_i}(t)$: This serves as the identifier of a vulnerability of asset $a_i$.
                \item $CVSS_{v_i}(t)$: This metric assigns a numeric value to the severity of vulnerability $v_i$. Each CVE has a corresponding CVSS value.
                \item $CAPEC_{w_i}(t)$: Each vulnerability (CVE) is a materialization of a weakness (CWE) $w_i$ that can be exploited using a concrete attack pattern (CAPEC). In many cases, each CWE has more than one CAPEC associated. Consequently, this field is a set that contains all the possible attack patterns that can exploit the vulnerability that is being analyzed.
            \end{itemize}

\begin{figure*}[!htb]
              \begin{center}
              \includegraphics[width=6.8in]{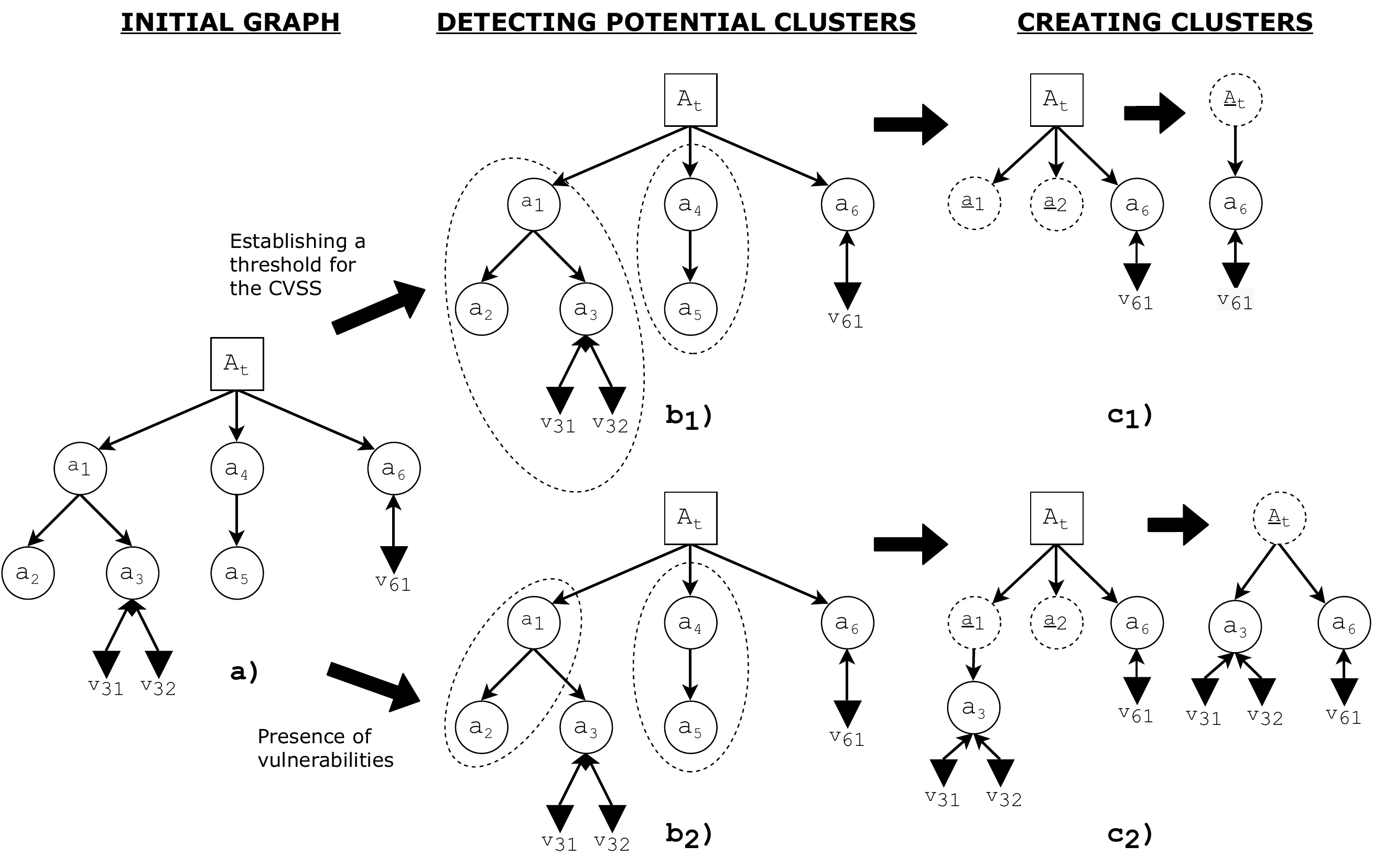}
              \caption{Creating clusters. Application of the two proposed criteria to the creation of clusters to simplify the graph: 
              (1) Establishing a threshold to select which vulnerability stays outside the cluster (upper side). 
              (2) Choosing the absence of vulnerability as the criterion to create clusters (lower side). The severity value (CVSS) for $v_{211}$ and $v_{212}$ is supposed to be lower than the establish threshold.}
              \label{fig:supernode_CriteriaapplicationExamples}
              \end{center}
            \end{figure*}

            \textbf{Clusters} \label{subsec:supernode} (collectively, set $G_{S}$) are a special type of node that summarizes and simplifies the information contained in a subgraph in an EDG. Fig. \ref{fig:supernode_CriteriaapplicationExamples} shows how the clusters work.
            
            % =======
            % FIG. XX
            % =======

            To identify each cluster, and to be able to recover the information that they summarize, each is characterized by the data that define each of the elements that they contain: $\{ CPE_{previous}, CPE_{current}, CWE_{a_i}(t) \}$, $\left( CVE_{a_i}(t), CVSS_{v_i}(t), \{ CAPEC_{w_i}(t) \right) \}$, and their dependencies.
    
            Two types of criteria can be used to create clusters and to simplify the obtained graph (Fig. \ref{fig:supernode_CriteriaapplicationExamples}:
            
            \begin{enumerate}
                \item \textbf{Absence of vulnerabilities}: Using this criterion, clusters will group all nodes that contain no associated vulnerabilities.
                \item \textbf{CVSS score below a certain threshold}: With this criterion, a threshold for the CVSS scores will be chosen. Nodes whose CVSS score is less than the defined threshold will be grouped into a cluster.
            \end{enumerate}

            % It is worth noticing that applying the second set of criteria (establishing a CVSS threshold) will always return a graph that is at least as simple or as complex as the one that would be obtained using the absence of vulnerability criterion. In the best case, the graph will be simpler. This is because both criteria treat assets with no vulnerabilities in the same way, so those will always be simplified. On the other hand, establishing a CVSS threshold allows the model for further simplifications.

        \subsubsection{Types of Edge}
            In the EDG model, edges plays a key role representing dependencies. Two types of edge can be identified:
            \begin{itemize}
                \item \textbf{Normal dependencies} relate two assets, or an asset and a vulnerability. They represent that the destination element depends on the source element. Collectively, they are known as set $G_D$.
                \item \textbf{Deprecated asset or patched vulnerability dependencies} indicate when an asset or a vulnerability is updated or patched. They represent that the destination element used to depend on the source element. Collectively, they are known as set $G_U$.
            \end{itemize}
    
            The possibility of representing old dependencies brings the opportunity to reflect the evolution of the SUT over time. When a new version of an asset is released, or a vulnerability is patched, the model will be updated. Their dependencies will change then from a normal dependency to a deprecated asset or vulnerability dependency to reflect that change.

        \vspace{0.1in}

        \subsubsection{Conditions of Application of EDGs}
            The EDG model is applicable to SUTs that meet the following set of criteria:
            \begin{itemize}
                \item \textbf{Software and hardware composition:} In our approach, the model is created by means of a white-box analysis. The absence or impossibility to perform a white-box analysis limits the ability to create an accurate model. Some knowledge about the internal structure and code is expected. This information is usually only known by the manufacturer of the component, unless the component is publicly available or open-source. It should be also possible to decompose the SUT into simpler assets to generate a relevant EDG.

                % \item \textbf{Operating System (OS):} The EDG model is capable of assessing SUTs either with or without OS. Moreover, depending on the SUT, OSs could be considered as a whole, being represented as a single asset, or they could be further decomposed into more manageable assets.
                
                \item \textbf{Existence of publicly known vulnerabilities:} The EDG model focuses on known vulnerabilities. This is not critical because many industrial components use commercial or open-source elements. The SUT must be composed of assets for which public information is available. If the majority of SUT assets are proprietary, or the SUT is an \textit{ad hoc} development that is never exposed, then the generated EDG will not evolve. Therefore, the analysis will not be relevant.
            \end{itemize}

        \subsubsection{Steps to Build the Model}
        \label{subsec:stepsToBuildTheModel}
            This section explains the process and algorithms that were used to build the corresponding EDG of a given SUT. The main scenarios that can be found are also described.
        
\begin{figure}[!htb]
                  \begin{center}
                  \includegraphics[width=2.5in]{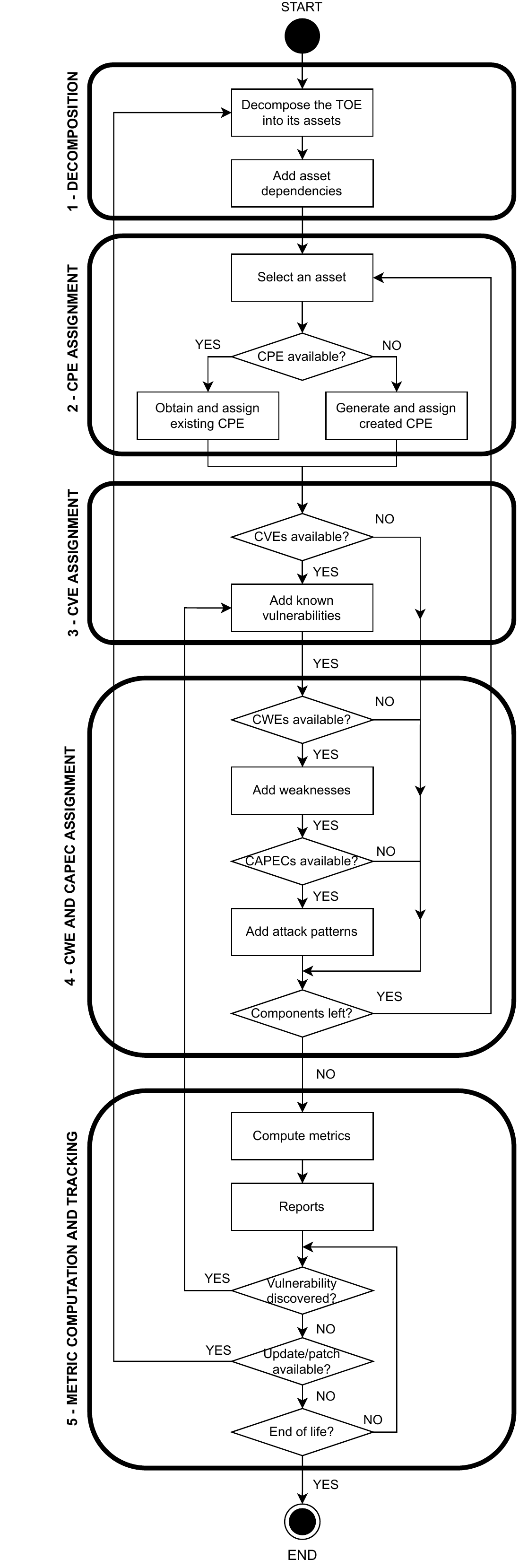}
                  \caption{Algorithm to generate the initial EDG of a give SUT.}
                  \label{fig:flowchart}
                  \end{center}
                \end{figure}

\begin{figure}[!htb]
                  \begin{center}
                  \includegraphics[width=2.9in]{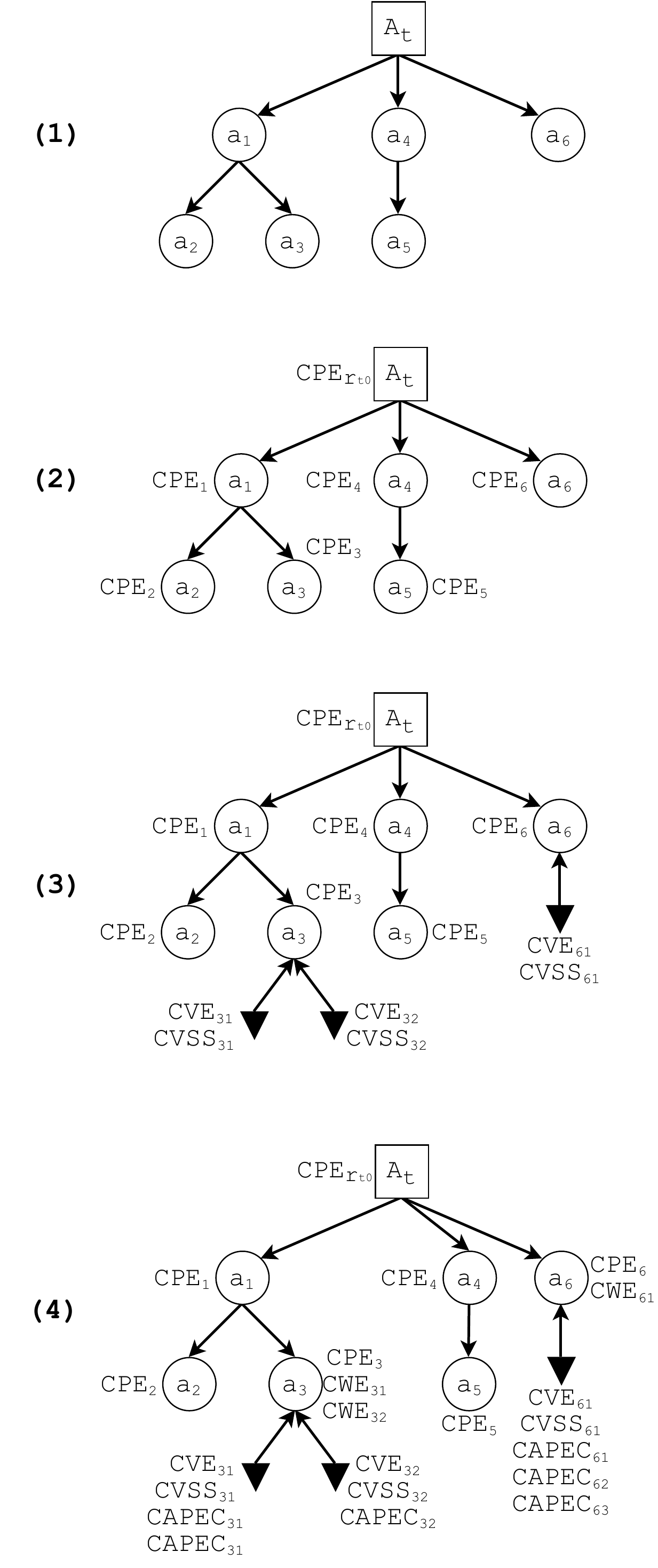}
                  \caption{Example of the process of building the EDG model of a given SUT $A$.}
                  \label{fig:steps_model}
                  \end{center}
                \end{figure}

            % Explicar cómo es el proceso de análisis (descomposición en activos, asignación de métrias, etc.)
            Before extracting useful information about the SUT, the directed graph associated with the SUT has to be built. This comprises several steps, which are described in the following paragraphs (see the flowchart in Fig. \ref{fig:flowchart}, and Fig. \ref{fig:steps_model}):

            \textbf{Step 1 --- Decompose the SUT into assets}.
                For the model to work properly, it relies on the SUT being able to be decomposed into assets. With this in mind, the first step involves obtaining the assets of the SUT, either software or hardware. In the CC, this process is called modular decomposition of the SUT~\cite{commonCriteriaGeneralModel}. Ideally, every asset should be represented in the decomposition process, but this is not compulsory for the model to work properly. Each one of the assets obtained in this step will be represented as an asset node. In this step, the dependencies among the obtained assets are also added as normal dependencies.
    
            \textbf{Step 2 --- Assign a CPE to each asset}.
                Once the assets and their dependencies have been identified, the next task is to assign the corresponding CPE identifier to each asset. If there is no publicly available information of a certain asset, and therefore, it does not have a CPE identifier, then it is always possible to generate one using the fields described in the CPE naming specification documents~\cite{CPE7695} for internal use in the model.
            
            \textbf{Step 3 --- Add known vulnerabilities to the assets}.
                In this step, the vulnerabilities ($CVE_{a_i}(t)$) of each asset are set. This is done by consulting public databases of known vulnerabilities~\cite{CVE1, NVD_nist} looking for existing vulnerabilities for each asset. When a vulnerability is found, it is added to the model of the SUT, including its dependencies. If there were no known vulnerabilities in an asset, then the asset would become the last leaf of its branch. In this step, the corresponding value of the CVSS of each vulnerability is also added to the model.
            
            \textbf{Step 4 --- Assign to each asset its weaknesses and possible CAPECs}.
                After the vulnerabilities, the corresponding weaknesses to each vulnerability ($CWE_{a_i}(t)$) are added, along with the corresponding attack patterns ($CAPEC_{w_i}(t)$) for each weakness. If there is no known vulnerability in an asset, then there will be no weaknesses.
                Meanwhile, it would be possible to have a known vulnerability in an asset, but no known weakness or attack pattern for that vulnerability.
                Finally, more than one CAPEC can be assigned to the same weakness. Consequently, it would be common to have a set of possible CAPECs that can be used to exploit the same weakness. It is worth noting that not all of them could be applied in every scenario.
                % ¿Podría no tener una debilidad asociada?
                % ¿Podría no tener un CAPEC asociado?
    
            \textbf{Step 5 --- Computing Metrics and tracking the SUT}.
                At this point, the EDG of the SUT is completed with all the public information that can be gathered. This last step is to calculate the metrics defined (for further information, see Section \ref{sec:securityMetrics}.), generating the corresponding reports, and tracking the state of the SUT for possible updates in the information of the model. This step is always triggered when the SUT is updated. This can imply that a new asset can appear, an old asset can disappear, an old vulnerability can be patched, or a new one can appear in the SUT. All of these scenarios will be reflected in the model as they arise during its life cycle.
                
                % =======
                % FIG. XX
                % =======
    
                % =======
                % FIG. XX
                % =======
                
                % Comentar que si no hay ninguna vulnerabilidad conocida, el diagrama será sólo un nodo.
                % After all these steps are applied, it might be the case that the resulting graph has only asset nodes and no vulnerabilities. This brings the opportunity to apply the concept of supernode described in Subsection \ref{subsec:supernode} to simplify the EDG, collapsing everything into a single supernode.
% =================================================================================
% === VI. SECURITY METRICS ========================================================
% =================================================================================
% \section{A Novel Methodology For Vulnerability Analysis}
    \subsection{Security Metrics}
    \label{sec:securityMetrics}
        % ¿Qué ventaja introduce el uso de métricas? ¿Qué permiten conseguir?
        The EDG model that was proposed in the previous sections is by itself capable of representing the internal structure of the SUT, and it can display it graphically for the user. This representation not only includes the internal assets of the SUT, but it also captures their relationships, existing vulnerabilities, and weaknesses. Moreover, assets, vulnerabilities, and weaknesses are easily identified using their corresponding CPE, CVE, and CWE values, respectively. All together, this constitutes a plethora of information that the model can use to improve the development and maintenance steps of the SUT, enhance its security, and track its status during its whole life cycle. Metrics are a great tool to integrate these features into the model.
        
        Metrics can serve as a tool to manage security, make decisions, and compare results over time. They can also be used to systematically improve the security level of an industrial component or to predict its security level in a future point in time.

        In this section, the basic definitions that serve as the foundation of the metrics are described. Then, the proposed metrics are introduced to complement the functionality of the EDG model. The main feature of these metrics is that they all depend on time as a variable, so it is possible to capture the actual state of the SUT, track its evolution over time, and compare the results.
        
        \subsubsection{Basic Definitions}
            In this section, the basic concepts on which the definitions of the metrics will be based are formalized.
        
            \begin{definition}
                The set of all possible weaknesses at a time $t$ is represented as $CWE(t)$, where $CWE(t) = \{cwe_1, ..., cwe_m\}$, and where $m$ is the total number of weaknesses at time t. This set contains the whole CWE database defined by MITRE~\cite{CWE0}.
            \end{definition}
    
            \begin{definition}
                The set of all of the possible vulnerabilities at a time $t$ is represented as $CVE(t)$, where $CVE(t) = \{cve_1, ..., cve_p\}$, and where $p$ is the total number of vulnerabilities. This set contains the whole CVE database defined by MITRE~\cite{CVE1}.
            \end{definition}
    
            \begin{definition}
                The set of all possible attack patterns at a time $t$ is represented as $CAPEC(t)$, where $CAPEC(t) = \{capec_1, ..., capec_q\}$, and where $q$ is the total number of attack patterns at time t. This set contains the whole CAPEC database defined by MITRE~\cite{CAPEC1}.
            \end{definition}
    
            \begin{definition}
                The set of weaknesses of an asset $a_i$ at a time $t$ is defined as $CWE_{a_i}(t) = \{cwe_j \vert cwe_j \text{ is in the asset } a_i \text{ at time } t \land cwe_j \in CWE(t) \land \forall k \neq j, cwe_j \neq cwe_k \}$. From this expression, the set of all the weaknesses of a particular asset throughout its life cycle is defined as $CWE_{a_i} = \bigcup_{t=1}^TCWE_{a_i}(t)$ where $\vert CWE_{a_i} \vert$ is the total number of non-repeated weaknesses in its entire life cycle.
            \end{definition}
    
            \begin{definition}
                The set of vulnerabilities of an asset $a_i$ at a time $t$ is defined as $CVE_{a_i}(t) = \{cve_j \vert cve_j \text{ is in the asset } a_i \text{ at time } t \land cve_j \in CVE(t) \}$. From this expression, the set of vulnerabilities of an asset throughout its entire life cycle is defined as $CVE_{a_i} = \bigcup_{t=1}^TCVE_{a_i}(t)$ where $\vert CVE_{a_i} \vert$ is the total number of vulnerabilities in its entire life cycle.
            \end{definition}
    
            \begin{definition}
                The set of weaknesses of a SUT $A$ with $n$ assets at a time $t$ is defined as:
                \begin{equation}
                    CWE_A(t) = \bigcup_{i=1}^{n}CWE_{a_i}(t)
                \end{equation}
            \end{definition}
    
            \begin{definition}
                The set of vulnerabilities of a SUT $A$ with $n$ assets at a time $t$ is defined as:
                \begin{equation}
                    CVE_A(t) = \bigcup_{i=1}^{n}CVE_{a_i}(t)
                \end{equation}
            \end{definition}

            \begin{definition}
                The set of vulnerabilities associated to the weakness $cwe_j$ and to the asset $a_i$ at a time $t$ is defined as:
                \begin{equation}
                    \begin{split}
                        CVE_{a_i | cwe_j}(t) = \{cve_k \vert cve_k \text{ associated to weakness }\\ cwe_j \text{ and to asset } a_i \text{ at time } t\}
                    \end{split}
                \end{equation}
            \end{definition}
            
            It is worth noting that CWE is used as a classification mechanism that differentiates CVEs by the type of vulnerability that they represent. A vulnerability will usually have only one associated weakness, and weaknesses can have one or more associated vulnerabilities~\cite{ARES}.
    
            \begin{definition}
                The partition $j$ of an asset $a_i$ at time $t$ conditioned by a weakness $cwe_k$ is defined as $CVE_{a_{i}\vert cwe_k}(t) = \{ cwe_l \vert cwe_l = cwe_k \land cwe_l \in CVE_{a_i}(t) \}$
            \end{definition}
    
            \begin{definition}
                The partition $j$ of the SUT $A$ at time $t$ conditioned by a weakness $cwe_k$ is defined as $CVE_{A\vert cwe_k}(t) = \{ cwe_l \vert cwe_l = cwe_k \land cwe_l \in CVE_{A}(t) \}$
            \end{definition}
    
            \begin{definition}
                The set of attack patterns associated to a weakness  $w_i$ at a time $t$ is defined as $CAPEC_{w_i}(t) = \{capec_j \vert capec_j \text{ can exploit weakness } w_i \text{ at time } t \land capec_j \in CAPEC(t) \}$.
            \end{definition}

            \begin{definition}
                $M = \{m_1, ..., m_r\}$ represents the set of metrics that are defined in this research work based on the EDG model, where $r$ is the total number of metrics. This set can be extended, defining more metrics according to the nature of the SUT.
            \end{definition}

        \subsubsection{Metrics}
            This section will describe the metrics that were defined based on the EDG model and the previous definitions. Although it might seem trivial, the most interesting feature of these metrics is that they all depend on time. Using time as an input variable for the computation of the metrics opens the opportunity to track results over time, compare them, and analyze the evolution of the status of the SUT. Furthermore, some metrics take advantage of time to generate an accumulated value, giving information about the life cycle of the SUT. Table \ref{tab:metrics} shows all of the proposed metrics, their definition, and their reference values.

\begin{table*}[!htb]
\centering
\caption{Proposed metrics for the model.}
\label{tab:metrics}
\resizebox{\textwidth}{!}{%
\begin{tabular}{@{}clp{0.5\linewidth}p{0.5\linewidth}@{}}
\hline
\multicolumn{2}{l}{METRIC} & DEFINITION & REFERENCE VALUE \\ \hline
\multirow{30}{*}{\rot{VULNERABILITIES}} & $M_0(A) = \frac{\vert CVE_{A}(t) \vert}{n(t)}$ & Arithmetic mean of vulnerabilities in the SUT $A$, where $n(t)$ is the number of assets in a SUT at a time $t$. $M_0$ shows how many vulnerabilities would be present in each asset if they were evenly distributed among the assets of the SUT. The result of $M_0$ can serve as a preliminary analysis of the SUT, related to the criticality of its state. & \begin{tabular}[c]{@{}l@{}}$M_0 < 1$: The number of vulnerabilities is lower than the number of assets.\\ $M_0 \geq 1$: Every asset has at least one vulnerability.\end{tabular} \\
 &  &  &  \\
 & $M_1( A, t ) = \vert CVE_A(t) \vert$ & Number of vulnerabilities in a SUT $A$ at time $t$. & Ideally, the values of $M_{2}$ should be zero (no vulnerability in $A$), but the lower the value of $M_2$, the better. \\
 &  &  &  \\
 & $M_2( A ) = \sum_{t=1}^{T}\vert CVE_A(t) \vert = \sum_{t=1}^{T} M_1( A, t )$ & Number of vulnerabilities in a SUT $A$ throughout its entire life cycle $T$. This metric computes the accumulated value of the number of vulnerabilities of a SUT throughout its entire life cycle. & The lower the value of $M_3$, the better. \\
 &  &  &  \\
 & $M_3( a_i, t ) = \vert CVE_{a_i}(t) \vert$ & Number of vulnerabilities in an asset $a_k$ at time $t$ The values of $M_3$ can be useful during a vulnerability analysis, or when performing a penetration test, to identify the asset with more vulnerabilities. & Ideally, the value of $M_5$ should be zero. \\
 &  &  &  \\
 & $M_4( a_k, t ) = \frac{\vert CVE_{a_k}(t) \vert}{\sum_{i=1}^{n} \vert CVE_{a_i}(t) \vert}$ & Relative frequency of vulnerabilities of the asset $a_k$ at a time $t$. & Ideally, the value of $M_6$ should be zero, or at least $M_6 \leq \frac{1}{n(t)}$, being $n(t)$ the number of assets in the SUT. This value can also be expressed as the percentage of vulnerabilities of asset $a_i$ respect to the total number of vulnerabilities in the SUT, $M_6( a_k, t ) = \frac{\vert CVE_{a_k}(t) \vert}{\sum_{i=1}^{n} \vert CVE_{a_i}(t) \vert} \text{·} 100$ \\
 &  &  &  \\
 & $M_5( a_i, cwe_j, t ) = \vert CVE_{a_i\vert cwe_j}(t)\vert$ & Multiplicity of weakness $cwe_j$ of the asset $a_i$ at a time $t$. This metric represents the number of times a weakness is present among the vulnerabilities of the asset $a_i$. This is possible because a vulnerability can have associated the same weakness as other vulnerabilities. & Ideally, the value of $M_7$ should be zero, or at least, $M_7 \leq \frac{\vert CVE_{A \vert cwe_{j}}(t) \vert}{n(t)}$, being $n(t)$ the number of assets in the SUT. The value of the metric could be further narrowed by assuming that $cwe_j$ will be present in all but one asset, so $M_7 \leq \frac{\vert CVE_{A \vert cwe_{j}}(t) \vert}{n(t) - 1}$ to be in acceptable values. \\
 &  &  &  \\
 & $M_6( A, cwe_j, t ) = \vert CVE_{A\vert cwe_j}(t)\vert$ & Multiplicity of weakness $cwe_j$ of the SUT $A$ at a time $t$. This metric represents the number of times a weakness is present among the vulnerabilities of the SUT $A$. & Ideally, the value of $M_8$ should be zero. \\ \hline
\multirow{7}{*}{\rot{WEAKNESSES}} &  &  &  \\
 & $M_7( A, t ) = \vert CWE_A(t) \vert$ & Number of weaknesses in a SUT $A$ at time $t$. & Ideally, the value of $M_{1}$ should be zero (no weakness in $A$), but the lower the value of $M_1$, the better. \\
 &  &  &  \\
 & $M_8( A ) = \sum_{t=1}^{T}\vert CWE_A(t) \vert = \sum_{t=1}^{T} M_7( A, t )$ & Number of weaknesses in a SUT $A$ throughout its entire life cycle $T$. This metric computes the accumulated value of weaknesses of a SUT throughout its entire life cycle. & The lower the value of $M_4$, the better. \\ \hline
\end{tabular}%
}
\end{table*}

            In addition to the metrics in Table \ref{tab:metrics}, the model allows the definition of other types of metrics according to the analysis to be performed, and the nature of the SUT (\textit{e.g.}, the vulnerability evolution function for SUT $A$ up to time $t$ for all vulnerabilities can be defined as the linear regression of the total number of vulnerabilities in each time $t$ for SUT $A$).

    \subsection{Properties}
    \label{sec:properties}
        Together, the EDG model and the defined metrics exhibit a series of characteristics that make them suitable for vulnerability assessment. These properties represent an advantage over the techniques reviewed in the state of the art, including automatic inference of root causes, spatial and temporal distribution of vulnerabilities, and prioritization of patching, which will be described in the following subsections.
        
        \subsubsection{Automatic Inference of Root Causes}
            Each CWE natively contains information that is directly related to the root cause of a vulnerability. From this information, new requirements and test cases can be proposed.

        \subsubsection{Spatial and Temporal Distribution of Vulnerabilities}
        \label{subsubsec:SpatialAndTemporalGranularModelling}
            The key feature of the proposed model is the addition of the temporal dimension in the analysis of vulnerabilities. This makes it possible to analyze the location of the vulnerabilities both in space (in which asset) and time (their recurrence), which allows us to track the state of the device throughout the whole life cycle. This approach also enables a further analysis of the SUT, by updating data in the model, such as new vulnerabilities that are found or new patches that are released.

\begin{figure*}[!htb]
              \begin{center}
              \includegraphics[width=6in]{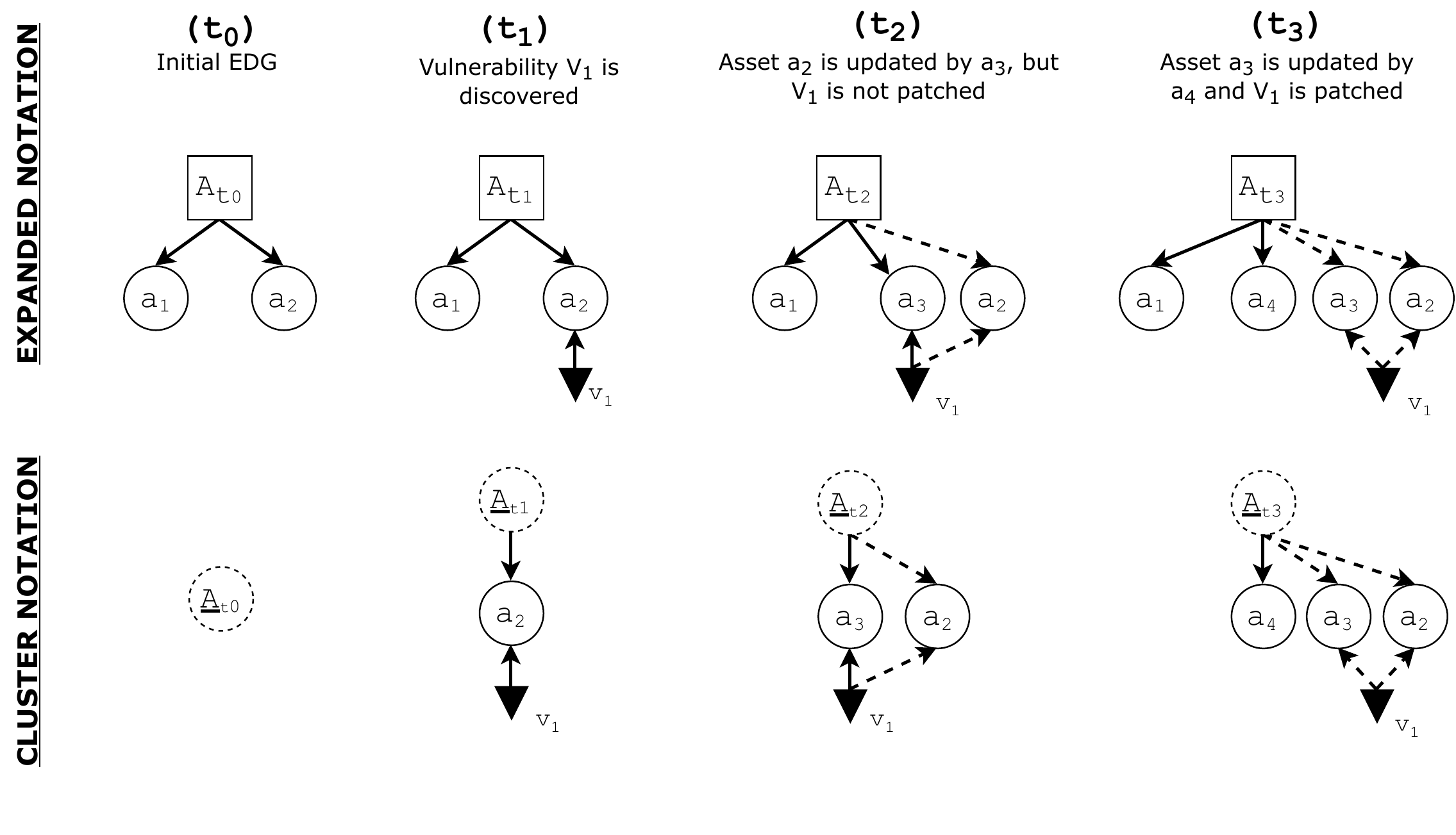}
              \caption{Representation of the temporal behavior in the graphical model using the two kinds of dependencies of the model. It is worth mentioning that these graphs could be further simplified by taking advantage of the cluster notation, as shown at the bottom of this figure.}
              \label{fig:time_model}
              \end{center}
            \end{figure*}

            Each time that a new vulnerability is found, or an asset is patched (\textit{i.e.}, via an update), the initial EDG is updated to reflect those changes. An example of this process can be seen in Fig. \ref{fig:time_model}.
    
            % =======
            % FIG. XX
            % =======
    
            At time $t_0$, the initial graph of the SUT $A$ is depicted in Fig. \ref{fig:time_model}. Because there is no vulnerability at that time, this graph can be simplified using the cluster notation, with just a cluster containing all assets. 
            At time $t_1$, a new vulnerability that affects the asset $a_2$ is discovered. 
            At time $t_2$, the asset $a_2$ is updated. This action creates a new version of asset $a_2$, asset $a_{3}$. Because the vulnerability was not corrected in the new update, both versions contain the vulnerability that was initially presented in asset $a_2$. 
            Finally, at time $t_3$, the asset $a_{3}$ is updated to its new version $a_{4}$, and the vulnerability is corrected.
            
            This approach enables a further analysis of the SUT, including updated data, according to new vulnerabilities that are found or new patches that are released.
       
            % Results obtained by the EDG model could lead to a massive graph with plenty of assets, vulnerabilities, and their corresponding dependencies. This can blur the visualization of the data, introducing difficulties for the interpretation of the results. The concept of \textit{supernode} was introduced in Subsection \ref{subsec:supernode} to correct this issue.

            % It is worth mentioning that the effectiveness of the creation of supernodes depends on the nature of the directed graph that is obtained. Figure \ref{fig:supernode_useCases} shows two possible scenarios. In the first one (Figure \ref{fig:supernode_useCases}a), all the vulnerabilities are located in the same asset, so it will be possible to simplify this graph applying the ``Absence of vulnerability'' criterion presented in Section \ref{subsec:supernode}. The complexity lies in the number of vulnerabilities, not in the number of assets. In the second one (Figure \ref{fig:supernode_useCases}b), as all vulnerabilities are distributed among all assets, only the ``CVSS below threshold'' criterion can be applied. Thus, the complexity lies in the number of assets and the distribution of vulnerabilities among them. This last scenario is not desirable, as it is an indicator of a transversal problem during the development of the SUT, as all assets have vulnerabilities.
    
            % % =======
            % % FIG. XX
            % % =======
            % \begin{figure}[!htb]
            %   \begin{center}
            %   \includegraphics[width=3in]{photo/supernode_cases.pdf}
            %   \caption{Possible scenarios for simplification of supernodes.}
            %   \label{fig:supernode_useCases}
            %   \end{center}
            % \end{figure}

        \subsubsection{Patching Policies Prioritization Support}
            The proposed model is not only able to include known vulnerabilities associated with an asset, but it also provides a relative importance sorting of vulnerabilities by CVSS. Relying on the resulting value, it is possible to assist in the vulnerability patching prioritization process. Furthermore, the presence of an existing exploit for a known vulnerability can be also be taken into account, when deciding which vulnerabilities need to be patched first. A high CVSS value combined with an available exploit for a given vulnerability is a priority when patching.
% =================================================================================
% === VIII. APPLICABILITY =========================================================
% =================================================================================
% \section{A Novel Methodology For Vulnerability Analysis}
    \subsection{Applicability in the Context of ISA/IEC 62443}
    \label{sec:applicability}
        In this section, the potential application of the proposed EDG model to the existing security standards is described. The proposed EDG model can be used isolated by itself, or in combination with other techniques that complement the analysis. In this sense, the EDG model can be used to enhance some task in the security evolution processes defined by security standards.

        % ISA/IEC 62443
        The ISA/IEC 62443-4-1 standard specifies 47 process requirements for the secure development of products used in industrial automation and control systems~\cite{IEC62443_4_1}. Thus, the EDG model was developed to enhance the execution of one of those requirements defined by the standard: the ``SVV-3: Vulnerability testing'' requirement, serving as a support for the execution of Practice 5 --- Security Verification and Validation testing. According to the SVV-3 requirement, both known and unknown vulnerability analysis has to be performed. The EDG model proposed in this research work is intended to support the identification of known vulnerabilities, their dependencies, and the possible consequences of their propagation, yielding the opportunity to analyze them systematically. Nevertheless, more requirements of the ISA/IEC 62443 can be mapped to one or more of the metrics defined in this research work. Using this relationship, it is possible to apply the EDG model to enhance the analysis and review of the following requirements:

        \subsubsection{Security Requirements - 2: Threat Model (SR-2)}
            ``A process shall be employed to ensure that all products have a threat model specific to the current development scope of the product. The threat model shall be reviewed and verified periodically''~\cite{IEC62443_4_1}. The proposed EDG model can serve as an abstraction of the threat model that has to be obtained. Moreover, the standard states that this threat model has to be reviewed periodically for updates. Given that the EDG of a given SUT evolves with every update, the threat model would be always up-to-date. Potential threats and their severity using the CVSS can also be analyzed with this proposal. Finally, these results can be used to enhance the risk assessment of the SUT.

        \subsubsection{Security Management - 13: Continuous Improvement (SM-13)}
            ``A process shall be employed for continuously improving the secure development life cycle''~\cite{IEC62443_4_1}. The EDG model can be used to identify recurrent issues in the development of an industrial component, due to its ability to track the state of a SUT over time. Consider the scenario where a piece of code contains an unknown vulnerability. For example, this code can implement a communication protocol, or the generation of a cryptographic key. If this piece of code is recurrently integrated in many type of devices, then when they are released to the market, the end users can identify that vulnerability and report it to the product supplier. The EDG can reflect the presence of that vulnerability. If an EDG is done for each type of device, then this problem can be detected beforehand. Using the CWE, the root problem can be detected. With this information, new training and corrective actions can be proposed to avoid this issue.

        \subsubsection{Specification of Security Requirements - 5: Security Requirements Review (SR-5)}
            ``A process shall be employed to ensure that security requirements are reviewed, updated, and approved''~\cite{IEC62443_4_1}. As before, taking advantage of the previous scenario, the information extracted from the generated EDG model can be used to propose new requirements or to update the existing requirements.

        \subsubsection{Security Verification and Validation Testing - 4: Penetration Testing (SVV-4)}
            ``A process shall be employed to identify and characterize security-related issues via tests that focus on discovering and exploiting security vulnerabilities in the product''~\cite{IEC62443_4_1}. The EDG model facilitates the identification of possible entry points to the SUT when carrying out a penetration test. In addition, existing attack patterns (CAPEC) and weaknesses (CWE) can serve as a starting point to discover unknown vulnerabilities and exploits.

        \subsubsection{Management of Security-related Issues - 3: Assessing Security-related issues (DM-3)}
            ``A process shall be employed for analyzing security-related issues in the product''~\cite{IEC62443_4_1}. When a new vulnerability is detected, end users will report it to the product suppliers. Then, the corresponding EDG model of that SUT will be updated to reflect that change. This information, in addition to that previously contained in the EDG, can be used to obtain the severity value of the discovered vulnerability using the CVSS. This also facilitates the identification of root causes, related security issues, or the impact.

        \begin{table}[!htb]
        \centering
        \caption{Mapping between the developed metrics and the requirements they refer in the ISA/IEC 62443.\\ SR (Security Requirements), SM (Security Management), SVV (Security Validation and Verification), DM (Management of Security-Related Issues).}
        \label{tab:mapping62443}
        \resizebox{0.48\textwidth}{!}{%
            \begin{tabular}{llllll}
            \hline
            \multicolumn{1}{c}{METRIC}                                                                 & \multicolumn{1}{c}{SR-2}           & \multicolumn{1}{c}{SR-5}           & \multicolumn{1}{c}{SM-13}          & \multicolumn{1}{c}{SVV-4}          & \multicolumn{1}{c}{DM-3}          \\ \hline
                                                                                                       &                                    &                                    &                                    &                                    &                                   \\
            $M_0(A) = \frac{\vert CVE_{A}(t) \vert}{n(t)}$                                             & \multicolumn{1}{c}{$\blacksquare$} & \multicolumn{1}{c}{$\blacksquare$} & \multicolumn{1}{c}{$\blacksquare$} & \multicolumn{1}{c}{$\blacksquare$} & \multicolumn{1}{c}{$\blacksquare$}\\
                                                                                                       &                                    &                                    &                                    &                                    &                                   \\
            $M_1( A, t ) = \vert CVE_A(t) \vert$                                                       & \multicolumn{1}{c}{$\blacksquare$} & \multicolumn{1}{c}{$\blacksquare$} & \multicolumn{1}{c}{$\blacksquare$} & \multicolumn{1}{c}{$\blacksquare$} & \multicolumn{1}{c}{$\blacksquare$}\\
                                                                                                       &                                    &                                    &                                    &                                    &                                   \\
			$M_2( A ) = \sum_{t=1}^{T}\vert CVE_A(t) \vert = \sum_{t=1}^{T} M_1( A, t )$               & \multicolumn{1}{c}{$\square$}      & \multicolumn{1}{c}{$\blacksquare$} & \multicolumn{1}{c}{$\blacksquare$} & \multicolumn{1}{c}{$\square$}      & \multicolumn{1}{c}{$\square$}     \\
                                                                                                       &                                    &                                    &                                    &                                    &                                   \\
            $M_3( A, t ) = \vert CVE_{a_i}(t) \vert$                                                   & \multicolumn{1}{c}{$\blacksquare$} & \multicolumn{1}{c}{$\blacksquare$} & \multicolumn{1}{c}{$\blacksquare$} & \multicolumn{1}{c}{$\blacksquare$} & \multicolumn{1}{c}{$\square$}     \\
                                                                                                       &                                    &                                    &                                    &                                    &                                   \\
			$M_4( a_k, t ) = \frac{\vert CVE_{a_k}(t) \vert}{\sum_{i=1}^{n} \vert CVE_{a_i}(t) \vert}$ & \multicolumn{1}{c}{$\square$}      & \multicolumn{1}{c}{$\blacksquare$} & \multicolumn{1}{c}{$\blacksquare$} & \multicolumn{1}{c}{$\square$}      & \multicolumn{1}{c}{$\square$}     \\
                                                                                                       &                                    &                                    &                                    &                                    &                                   \\
            $M_5( a_i, cwe_j, t ) = \vert CVE_{a_i\vert cwe_j}(t)\vert$                                & \multicolumn{1}{c}{$\blacksquare$} & \multicolumn{1}{c}{$\blacksquare$} & \multicolumn{1}{c}{$\blacksquare$} & \multicolumn{1}{c}{$\blacksquare$} & \multicolumn{1}{c}{$\square$}     \\
                                                                                                       &                                    &                                    &                                    &                                    &                                   \\
			$M_6( A, cwe_j, t ) = \vert CVE_{A\vert cwe_j}(t)\vert$                                  & \multicolumn{1}{c}{$\blacksquare$} & \multicolumn{1}{c}{$\blacksquare$} & \multicolumn{1}{c}{$\square$}      & \multicolumn{1}{c}{$\blacksquare$} & \multicolumn{1}{c}{$\blacksquare$}\\
                                                                                                       &                                    &                                    &                                    &                                    &                                   \\
            $M_7( A, t ) = \vert CWE_A(t) \vert$                                                       & \multicolumn{1}{c}{$\blacksquare$} & \multicolumn{1}{c}{$\square$}      & \multicolumn{1}{c}{$\square$}      & \multicolumn{1}{c}{$\blacksquare$} & \multicolumn{1}{c}{$\blacksquare$}\\
                                                                                                       &                                    &                                    &                                    &                                    &                                   \\
            $M_8( A ) = \bigcup_{t=1}^{T}\vert CWE_A(t) \vert = \bigcup_{t=1}^{T} M_7( A, t )$         & \multicolumn{1}{c}{$\square$}      & \multicolumn{1}{c}{$\blacksquare$} & \multicolumn{1}{c}{$\blacksquare$} & \multicolumn{1}{c}{$\square$}      & \multicolumn{1}{c}{$\square$}     \\
                                                                                                       &                                    &                                    &                                    &                                    &                                   \\ \hline           
            \end{tabular}%
            }
        \end{table}

        \vspace{0.2in}

        Finally, the ISA/IEC 62443-4-2 document defines four types of components of an IACS (\textit{i.e.}, software applications, embedded devices, host devices, network devices)~\cite{IEC62443_4_2}. The proposed model is capable of representing the inherent complexity of each of them.
% =================================================================================
% === VIII. Real Use Case Example =================================================
% =================================================================================
\section{Real Use Case}
\label{sec:usecase}
    In this section, the EDG model and the proposed metrics will be applied to perform a vulnerability assessment of the OpenPLC project, which will be the SUT. In the subsections, we will assess the three available versions of the OpenPLC project. For each, the EDG model will be obtained, and the proposed metrics will be applied to draw conclusions about the vulnerability status of each version.
    
    OpenPLC is the first functional standardized open source Programmable Logic Controller (PLC), both in software and hardware~\cite{OpenPLCProject}. It was mainly created for research purposes in the areas of industrial and home automation, Internet of Things (IoT), and SCADA. Given that it is the only controller that provides its entire source code, it represents an engaging low-cost industrial solution --- not only for academic research but also for real-world automation~\cite{Alves_openPLC_2014, Alves_OpenPLC_2018}.

    \subsection{Structure of OpenPLC}
        The OpenPLC project consists of three parts:
        \begin{enumerate}
            \item \textbf{Runtime:} It is the software that plays the same role as the firmware in a traditional PLC. It executes the control program. The runtime can be installed in a variety of embedded platforms, such as the Raspberry Pi, and in Operating Systems (OSs) such as Windows or Linux. Industrial Modbus slave devices can be attached to expand the number of inputs and outputs. % This is particularly useful for systems that do not have any input or output at all, \textit{e.g.}, when OpenPLC is running on Windows or Linux.
            \item \textbf{Editor:} An application that runs on a Windows or Linux OS that is used to write and compile the control programs that will be later executed by the runtime.
            \item \textbf{HMI Builder:} This software is to create web-based animations that will reflect the state of the process, in the same manner as a traditional HMI.
        \end{enumerate}
        When installed, the OpenPLC runtime executes a built-in webserver that allows OpenPLC to be configured and new programs for it to run to be uploaded.

    \subsection{Setup Through the Analysis}
        The OpenPLC project is constituted by three different versions~\cite{OpenPLCv1, OpenPLCv2, OpenPLCv3} as can be seen in Table \ref{tab:openPLC_versions}. For this research work, Ubuntu Linux was selected as the platform to install the OpenPLC runtime. Ubuntu Linux provides comprehensive documentation, previous versions are accessible, and software dependencies can easily be obtained.

        To make the analysis of OpenPLC fair, a contemporary operating system was selected for each of the OpenPLC versions, according to the version of Ubuntu that was available at the release time of each OpenPLC version (see Table \ref{tab:openPLC_versions}). The Long Term Support (LTS) version was chosen, given that the industry tends to work with the most stable version available of any software and security updates are provided for a longer time.

        \begin{table}[!htb]
        \centering
        \caption{Versions and release dates of OpenPLC and the available Ubuntu Linux LTS at that time for each date.}
        \label{tab:openPLC_versions}
        \resizebox{3.4in}{!}{%
            \begin{tabular}{@{}llll@{}}
                \toprule
                VERSION & RELEASE & OS & OS RELEASE \\ \midrule
                OpenPLC V1 & 2016/02/05 & Ubuntu 14.04 LTS & 2014/04/17 \\
                OpenPLC V2 & 2016/05/13 & Ubuntu 16.04 LTS & 2016/04/21 \\
                OpenPLC V3 & 2018/06/14 & Ubuntu 18.04 LTS & 2018/04/26 \\ \bottomrule
            \end{tabular}%
        }
        \end{table}
        
        The scenario used for the analysis consists of OpenPLC installed on Ubuntu Linux in a virtual machine, following the OSs shown in Table \ref{tab:openPLC_versions}.

    \subsection{Steps of the Analysis}
        The EDG model of OpenPLC was built following the steps described in Section \ref{subsec:stepsToBuildTheModel} (see flowchart in Fig. \ref{fig:flowchart}). It is worth noting that these steps are followed for each one of the three versions available of OpenPLC, so an EDG is generated for each one. Obtaining the EDG for each version of OpenPLC will give information about the evolution of vulnerabilities over time.

        For the sake of clarity and ease of analysis, the three obtained dependency graphs for each OpenPLC version were not merged into a single diagram. Fig. \ref{fig:openplcV1}, Fig. \ref{fig:openplcV2}, and Fig. \ref{fig:openplcV3}, show the obtained EDG for versions V1, V2, and V3, respectively. In reality, these three diagrams would be the result of applying this method over time, updating the graph each time that a new vulnerability is discovered or a new patch/update is issued.

    \subsection{Analysis}
        In this stage, the analysis of the SUT is performed based on the generated EDG and the value of the computed metrics. This process can be structured into three main steps, as follows:
        
        \begin{enumerate}
            \item \textbf{Analysis of the induced EDG model:} This step involves the analysis of the obtained directed-graph model. The structure, assets and dependencies are the focus of this first step.
            \item \textbf{Vulnerability analysis:} Vulnerability number, distribution, and severity are analyzed in this step, which is supported by metrics. A proposal for vulnerability prioritization is also proposed.
            \item \textbf{Root causes analysis (weaknesses):} Finally, the root cause of each vulnerability if found (related to the associated weakness). In this step, new requirements, test cases, and training activities are proposed based on the results of the analysis.
        \end{enumerate}

        For each of the previously described steps, the analysis is done in both the spatial and temporal dimensions:

        \begin{enumerate}
            \item \textbf{Spatial Dimension:} This kind of analysis focuses on the distribution of vulnerabilities among the assets at a time $t$. In this example, this corresponds to independently analyzing each obtained EDG, because each one represents an instant in time (a different version of OpenPLC).
            \item \textbf{Temporal Dimension:} This kind of analysis focuses on the evolution and distribution of vulnerabilities over time. In this example, this corresponds to analyzing the changes in the number of assets and vulnerabilities, and their distribution over all three EDGs (over all three versions of OpenPLC).
        \end{enumerate}

        \subsubsection{Analysis of the Induced EDG Model}
            OpenPLC V1 is analyzed in this subsection, focusing on the internal structure and dependencies among the assets.
            
\begin{figure*}[!htb]
              \begin{center}
              \includegraphics[width=6.8in]{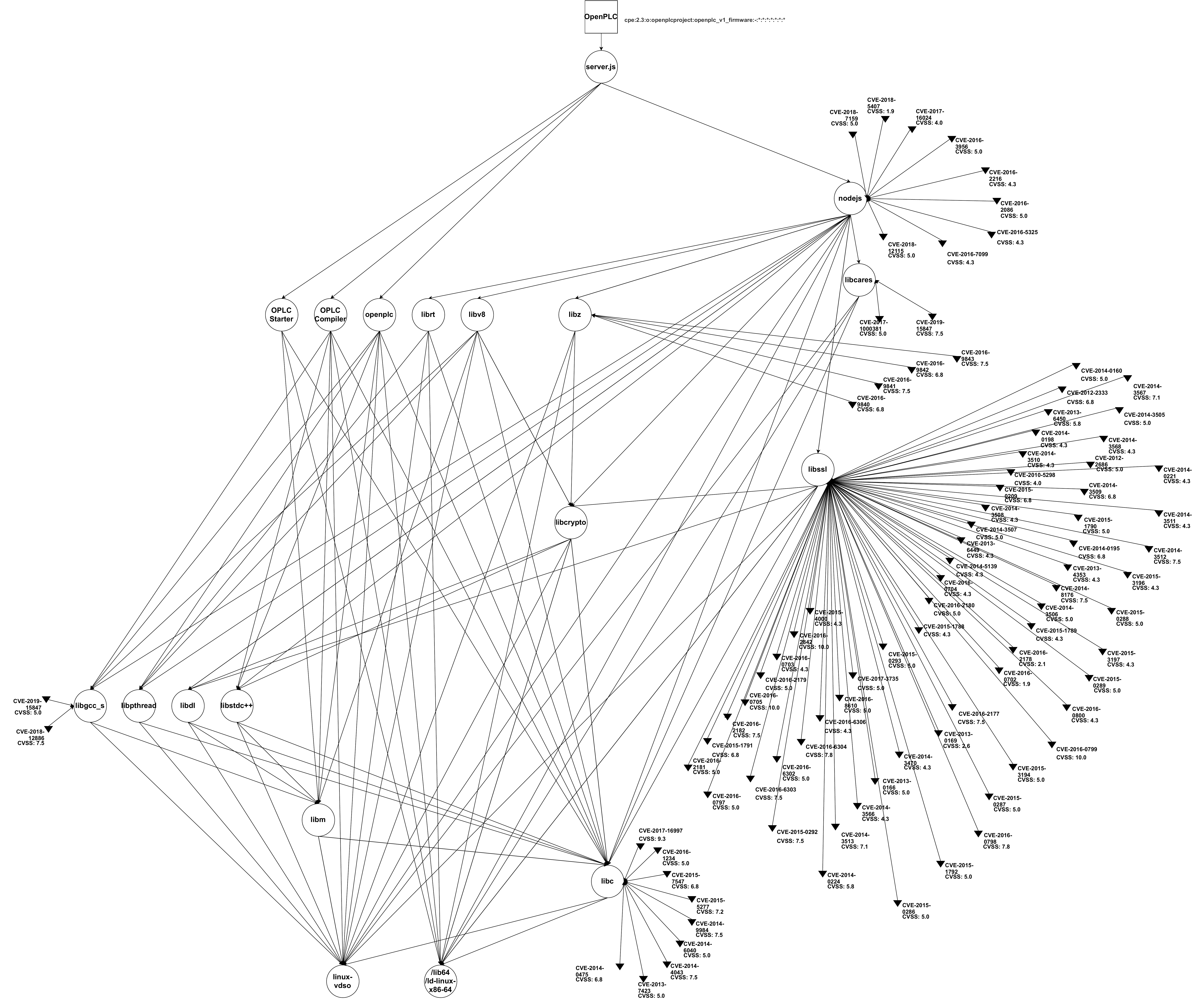}
              \caption{EDG for OpenPLC V1. Notice that, for simplicity, CWE, and CAPEC values are omitted, and only the CPE identifier of the SUT is shown.}
              \label{fig:openplcV1}
              \end{center}
            \end{figure*}

            The first result of the EDG model is thegraph obtained for OpenPLC V1 (Fig. \ref{fig:openplcV1}). From the \textbf{spatial dimension} point of view, assets depend on a main service, \texttt{server.js}, based in Java. This web server offers a web GUI for the user to configure, start, and stop the PLC execution. Below this level, the main components of OpenPLC can be seen: 
                OPLC Starter (responsible to start the OpenPLC and constantly monitor if it is running or not), 
                OPLC Compiler (compiler from ladder logic to ANSI C code), and 
                openplc (initialization procedures for the hardware, network and the main loop). 
            The other assets are dynamic libraries of the system, such as \texttt{libstdc++}, \texttt{libm}, texttt{libc} (for C programming), and texttt{libssl} (C library for SSL and TLS).

            Moreover, the \texttt{libc} library is a wrapper around the system calls of the Linux kernel, which provides and defines system calls and other basic functions. Thus, it is expected that all of the identified assets depend on this library in every version of OpenPLC. Fig. \ref{fig:openplcV1} shows that this is indeed the case.

            % =======
            % FIG. XX
            % =======

\begin{figure*}[!htb]
              \begin{center}
              \includegraphics[width=6.8in]{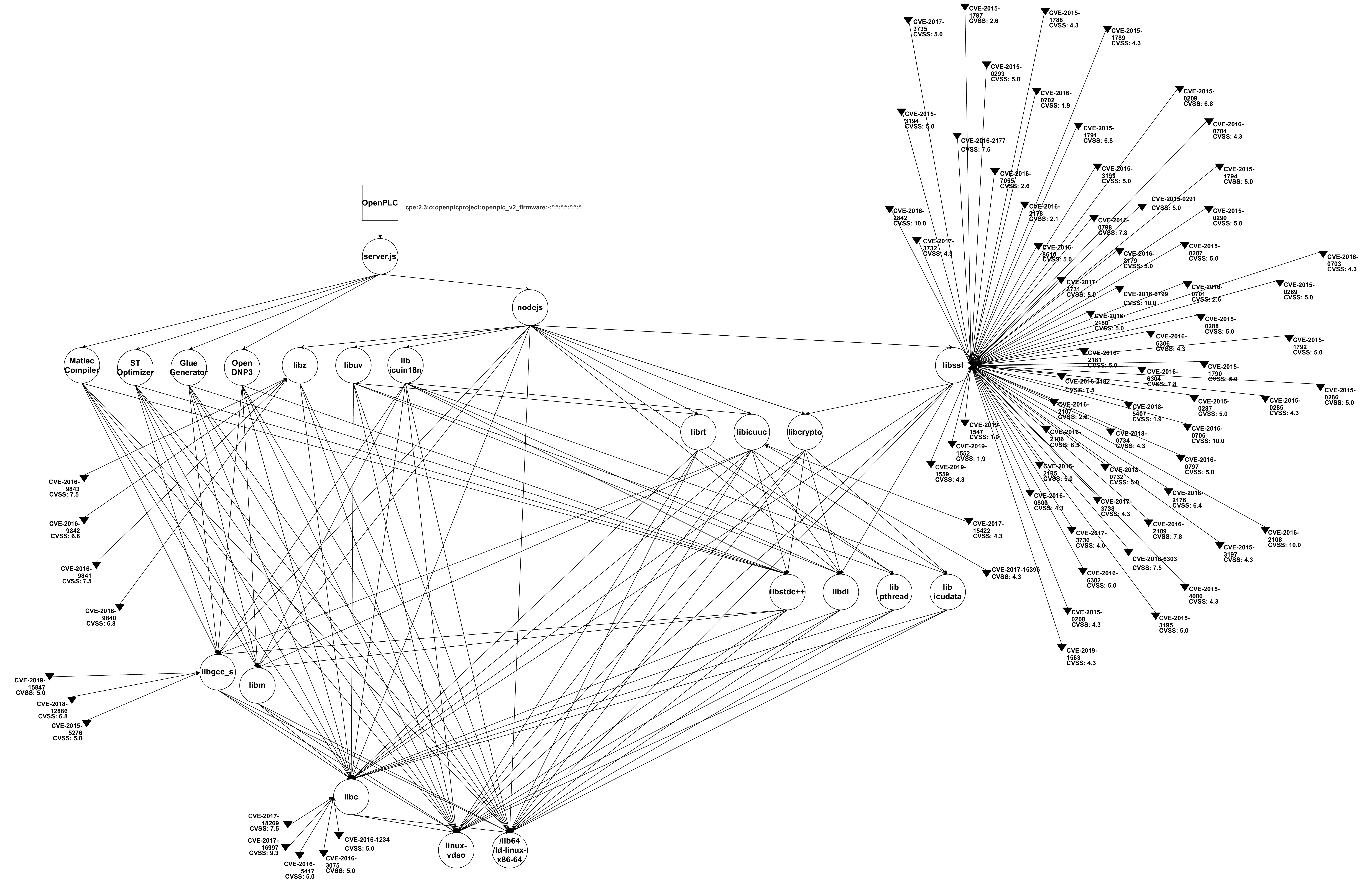}
              \caption{EDG for OpenPLC V2. Note that for the sake of simplicity, CWE, and CAPEC values are omitted and only the CPE identifier of the SUT is shown.}
              \label{fig:openplcV2}
              \end{center}
            \end{figure*}

            From the \textbf{temporal dimension} point of view, OpenPLC V2 has to be analyzed in comparison to the previous version of OpenPLC to find changes in the structure and the number of assets. Fig. \ref{fig:openplcV2} shows the EDG for OpenPLC V2. Comparing both OpenPLC V1 and OpenPLC v2, it can be noted that their structure is very similar. In OpenPLC V2, new assets were introduced to provide new features to the project:
                Matiec compiler (which is an open-source compiler for programming languages defined in the IEC 61131-3 standard), 
                ST optimizer (which is responsible for the optimization process after the initial compilation from OpenPLC Editor), 
                Glue Generator (which is responsible for gluing the variables from the IEC program to the OpenPLC memory pointers), and 
                OpenDNP3 (which is an implementation of the DNP3 protocol stack written in C++11).
            
            In OpenPLC V2, and in comparison with OpenPLC V1, the OPLC compiler has disappeared and has been substituted by the Matiec Compiler, which supports all programming languages defined in the IEC 61131-3 standard. Moreover, the ST Optimizer and the Glue Generator were added to support the compilation process. Finally, the OpenDNP3 library has been added for high-performance applications, such as many concurrent TCP sessions.

            % =======
            % FIG. XX
            % =======

\begin{figure*}[!htb]
              \begin{center}
              \includegraphics[width=6.8in]{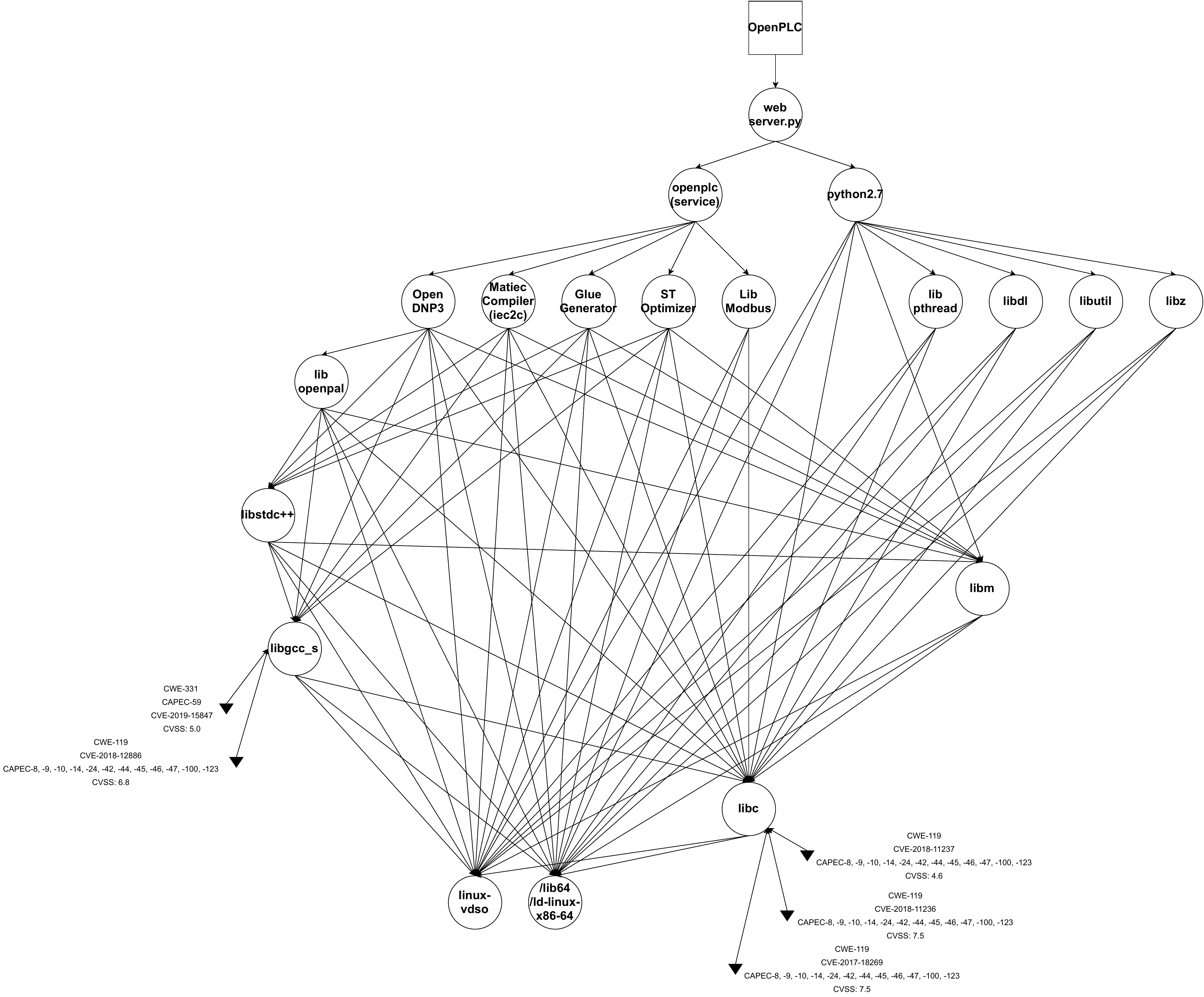}
              \caption{EDG for OpenPLC V3. Note that only the CPE of SUT is shown for the sake of simplicity.}
              \label{fig:openplcV3}
              \end{center}
            \end{figure*}

            Finally, and keeping the analysis in the temporal dimension, OpenPLC V3 can be compared to its ancestor, OpenPLC V2. Fig. \ref{fig:openplcV3} shows the EDG for OpenPLC V3. This last version of the project is the simplest. Now, the Java-based web server has been replaced by a Python-based web server, \textit{webserver.py}. The main components of this version are: 
                Matiec compiler, 
                ST optimizer, 
                Glue Generator, 
                OpenDNP3, and 
                LibModbus. 
            
            % =======
            % FIG. XX
            % =======

        \subsubsection{Vulnerability Analysis}
            In this step, all three available versions of the OpenPLC project are analyzed from a vulnerability perspective, both in the spatial and temporal dimensions. The goal here is to use the defined metrics as a support of the analysis, obtaining the number of vulnerabilities, their distribution, and their severity score. Table \ref{tab:metricsOpenPLC} shows the values of the proposed metrics for all three versions of OpenPLC. Finally, a ranked prioritization by severity for the vulnerabilities of each version is proposed.

\begin{table*}[!htb]
\centering
\caption{Metric values for each asset and version of OpenPLC. Notice that ``CWE-NULL'' refers to a void value of CWE for a certain CVE value.}
\label{tab:metricsOpenPLC}
\resizebox{\textwidth}{!}{%
\begin{tabular}{c|lr|cccccccccccccccc}
\hline
\multicolumn{3}{c|}{\multirow{2}{*}{METRIC}} & \multicolumn{7}{c|}{OpenPLC V1} & \multicolumn{6}{c|}{OpenPLC V2} & \multicolumn{3}{c}{OpenPLC V3} \\
\multicolumn{3}{c|}{} & \multicolumn{1}{l}{libgcc\_s} & \multicolumn{1}{l}{libc} & \multicolumn{1}{l}{libz} & \multicolumn{1}{l}{libcares} & \multicolumn{1}{l}{nodejs} & \multicolumn{1}{l}{libssl} & \multicolumn{1}{l|}{others} & \multicolumn{1}{l}{libgcc\_s} & \multicolumn{1}{l}{libc} & \multicolumn{1}{l}{libz} & \multicolumn{1}{l}{libicuuc} & \multicolumn{1}{l}{libssl} & \multicolumn{1}{l|}{others} & \multicolumn{1}{l}{libgcc\_s} & \multicolumn{1}{l}{libc} & \multicolumn{1}{l}{others} \\ \hline
\multirow{7}{*}{\rot{vulnerabilities}} & \multicolumn{2}{l|}{$n(t)$} & \multicolumn{7}{c|}{19} & \multicolumn{6}{c|}{22} & \multicolumn{3}{c}{19} \\ \cline{2-19} 
 & \multicolumn{2}{l|}{$M_0(A) = \frac{\vert CVE_{A}(t) \vert}{n(t)}$} & \multicolumn{7}{c|}{4.79} & \multicolumn{6}{c|}{3.50} & \multicolumn{3}{c}{0.26} \\ \cline{2-19} 
 & \multicolumn{2}{l|}{$M_1( A, t ) = \vert CVE_A(t) \vert$} & \multicolumn{7}{c|}{91} & \multicolumn{6}{c|}{77} & \multicolumn{3}{c}{5} \\ \cline{2-19} 
 & \multicolumn{2}{l|}{$M_2( A ) = \sum_{t=1}^{T}\vert CVE_A(t) \vert = \sum_{t=1}^{T} M_1( A, t )$} & \multicolumn{16}{c}{173} \\ \cline{2-19} 
 & \multicolumn{2}{l|}{$M_3(A) = \vert CVE_{a_i}(t) \vert$} & 2 & 9 & 4 & 2 & 9 & 65 & \multicolumn{1}{c|}{0} & 3 & 5 & 4 & 2 & 63 & \multicolumn{1}{c|}{0} & 2 & 3 & 0 \\ \cline{2-19} 
 & \multicolumn{2}{l|}{$M_4( a_k, t ) = \frac{\vert CVE_{a_k}(t) \vert}{\sum_{i=1}^{n} \vert CVE_{a_i}(t) \vert}$} & 0.02 & 0.10 & 0.04 & 0.02 & 0.10 & 0.71 & \multicolumn{1}{c|}{0.00} & 0.04 & 0.06 & 0.05 & 0.03 & 0.82 & \multicolumn{1}{c|}{0.00} & 0.40 & 0.60 & 0.00 \\ \hline
\multirow{46}{*}{\rot{weaknesses}} & \multirow{22}{*}{$M_5( a_i, cwe_j, t ) = \vert CVE_{a_i\vert cwe_j}(t)\vert$} & CWE-17 & - & 1 & - & - & - & 2 & \multicolumn{1}{c|}{-} & - & - & - & - & 3 & \multicolumn{1}{c|}{-} & - & - & - \\
 &  & CWE-19 & - & - & - & - & 1 & - & \multicolumn{1}{c|}{-} & - & - & - & - & - & \multicolumn{1}{c|}{-} & - & - & - \\
 &  & CWE-20 & - & - & - & - & 3 & 5 & \multicolumn{1}{c|}{-} & - & - & - & - & 3 & \multicolumn{1}{c|}{-} & - & - & - \\
 &  & CWE-22 & - & 1 & - & - & - & - & \multicolumn{1}{c|}{-} & - & - & - & - & - & \multicolumn{1}{c|}{-} & - & - & - \\
 &  & CWE-94 & - & 1 & - & - & - & - & \multicolumn{1}{c|}{-} & - & - & - & - & - & \multicolumn{1}{c|}{-} & - & - & - \\
 &  & CWE-113 & - & - & - & - & 1 & - & \multicolumn{1}{c|}{-} & - & - & - & - & - & \multicolumn{1}{c|}{-} & - & - & - \\
 &  & CWE-119 & 1 & 5 & - & - & - & 9 & \multicolumn{1}{c|}{-} & 1 & 3 & - & 1 & 6 & \multicolumn{1}{c|}{-} & 1 & 3 & - \\
 &  & CWE-125 & - & - & - & - & - & 2 & \multicolumn{1}{c|}{-} & - & - & - & - & 3 & \multicolumn{1}{c|}{-} & - & - & - \\
 &  & CWE-189 & - & - & 4 & - & - & 2 & \multicolumn{1}{c|}{-} & - & - & 4 & - & 4 & \multicolumn{1}{c|}{-} & - & - & - \\
 &  & CWE-190 & - & - & - & - & - & 1 & \multicolumn{1}{c|}{-} & - & - & - & 1 & 1 & \multicolumn{1}{c|}{-} & - & - & - \\
 &  & CWE-200 & - & - & - & 1 & 3 & 5 & \multicolumn{1}{c|}{-} & 1 & - & - & - & 12 & \multicolumn{1}{c|}{-} & - & - & - \\
 &  & CWE-295 & - & - & - & - & - & - & \multicolumn{1}{c|}{-} & - & - & - & - & 1 & \multicolumn{1}{c|}{-} & - & - & - \\
 &  & CWE-310 & - & - & - & - & - & 12 & \multicolumn{1}{c|}{-} & - & - & - & - & 5 & \multicolumn{1}{c|}{-} & - & - & - \\
 &  & CWE-311 & - & - & - & - & - & - & \multicolumn{1}{c|}{-} & - & - & - & - & 2 & \multicolumn{1}{c|}{-} & - & - & - \\
 &  & CWE-320 & - & - & - & - & - & - & \multicolumn{1}{c|}{-} & - & - & - & - & 3 & \multicolumn{1}{c|}{-} & - & - & - \\
 &  & CWE-331 & 1 & - & - & - & - & - & \multicolumn{1}{c|}{-} & 1 & - & - & - & - & \multicolumn{1}{c|}{-} & 1 & - & - \\
 &  & CWE-362 & - & - & - & - & - & 4 & \multicolumn{1}{c|}{-} & - & - & - & - & 1 & \multicolumn{1}{c|}{-} & - & - & - \\
 &  & CWE-399 & - & - & - & - & - & 8 & \multicolumn{1}{c|}{-} & - & 1 & - & - & 6 & \multicolumn{1}{c|}{-} & - & - & - \\
 &  & CWE-400 & - & - & - & - & - & 1 & \multicolumn{1}{c|}{-} & - & - & - & - & 1 & \multicolumn{1}{c|}{-} & - & - & - \\
 &  & CWE-426 & - & 1 & - & - & - & - & \multicolumn{1}{c|}{-} & - & 1 & - & - & - & \multicolumn{1}{c|}{-} & - & - & - \\
 &  & CWE-787 & - & - & - & 1 & 1 & 2 & \multicolumn{1}{c|}{-} & - & - & - & - & 2 & \multicolumn{1}{c|}{-} & - & - & - \\
 &  & CWE-NULL & - & - & - & - & - & 12 & \multicolumn{1}{c|}{-} & - & - & - & - & 10 & \multicolumn{1}{c|}{-} & - & - & - \\ \cline{2-19} 
 & \multirow{22}{*}{$M_6( A, cwe_j, t ) = \vert CVE_{A\vert cwe_j}(t)\vert$} & CWE-17 & \multicolumn{7}{c|}{3} & \multicolumn{6}{c|}{3} & \multicolumn{3}{c}{-} \\
 &  & CWE-19 & \multicolumn{7}{c|}{1} & \multicolumn{6}{c|}{-} & \multicolumn{3}{c}{-} \\
 &  & CWE-20 & \multicolumn{7}{c|}{8} & \multicolumn{6}{c|}{3} & \multicolumn{3}{c}{-} \\
 &  & CWE-22 & \multicolumn{7}{c|}{1} & \multicolumn{6}{c|}{-} & \multicolumn{3}{c}{-} \\
 &  & CWE-94 & \multicolumn{7}{c|}{1} & \multicolumn{6}{c|}{-} & \multicolumn{3}{c}{-} \\
 &  & CWE-113 & \multicolumn{7}{c|}{1} & \multicolumn{6}{c|}{-} & \multicolumn{3}{c}{-} \\
 &  & CWE-119 & \multicolumn{7}{c|}{15} & \multicolumn{6}{c|}{11} & \multicolumn{3}{c}{4} \\
 &  & CWE-125 & \multicolumn{7}{c|}{2} & \multicolumn{6}{c|}{3} & \multicolumn{3}{c}{-} \\
 &  & CWE-189 & \multicolumn{7}{c|}{6} & \multicolumn{6}{c|}{8} & \multicolumn{3}{c}{-} \\
 &  & CWE-190 & \multicolumn{7}{c|}{1} & \multicolumn{6}{c|}{2} & \multicolumn{3}{c}{-} \\
 &  & CWE-200 & \multicolumn{7}{c|}{9} & \multicolumn{6}{c|}{13} & \multicolumn{3}{c}{-} \\
 &  & CWE-295 & \multicolumn{7}{c|}{-} & \multicolumn{6}{c|}{1} & \multicolumn{3}{c}{-} \\
 &  & CWE-310 & \multicolumn{7}{c|}{12} & \multicolumn{6}{c|}{5} & \multicolumn{3}{c}{-} \\
 &  & CWE-311 & \multicolumn{7}{c|}{-} & \multicolumn{6}{c|}{2} & \multicolumn{3}{c}{-} \\
 &  & CWE-320 & \multicolumn{7}{c|}{-} & \multicolumn{6}{c|}{3} & \multicolumn{3}{c}{-} \\
 &  & CWE-331 & \multicolumn{7}{c|}{1} & \multicolumn{6}{c|}{1} & \multicolumn{3}{c}{1} \\
 &  & CWE-362 & \multicolumn{7}{c|}{4} & \multicolumn{6}{c|}{1} & \multicolumn{3}{c}{-} \\
 &  & CWE-399 & \multicolumn{7}{c|}{8} & \multicolumn{6}{c|}{7} & \multicolumn{3}{c}{-} \\
 &  & CWE-400 & \multicolumn{7}{c|}{1} & \multicolumn{6}{c|}{1} & \multicolumn{3}{c}{-} \\
 &  & CWE-426 & \multicolumn{7}{c|}{1} & \multicolumn{6}{c|}{1} & \multicolumn{3}{c}{-} \\
 &  & CWE-787 & \multicolumn{7}{c|}{4} & \multicolumn{6}{c|}{2} & \multicolumn{3}{c}{-} \\
 &  & CWE-NULL & \multicolumn{7}{c|}{12} & \multicolumn{6}{c|}{10} & \multicolumn{3}{c}{-} \\ \cline{2-19} 
 & \multicolumn{2}{l|}{$M_7( A, t ) = \vert CWE_A(t) \vert$} & \multicolumn{7}{c|}{19} & \multicolumn{6}{c|}{18} & \multicolumn{3}{c}{2} \\ \cline{2-19} 
 & \multicolumn{2}{l|}{$M_8( A ) = \bigcup_{t=1}^{T}\vert CWE_A(t) \vert = \bigcup_{t=1}^{T} M_7( A, t )$} & \multicolumn{16}{c}{22} \\ \hline
\end{tabular}%
}
\end{table*}

            From the \textbf{spatial dimension} perspective, just by observing the generated EDG for OpenPLC V1 (Fig. \ref{fig:openplcV1}), it can be stated that:
            
            \begin{itemize}
                \item Most of the vulnerabilities are present in third-party open-source components, such as \texttt{libssl} ($M_3 = 65$), and \texttt{node-js} ($M_3 = 9$), having \texttt{libssl} the $M_4 = 71.4 \%$ of the vulnerabilities in this version. \textit{A priori}, just by looking the number of vulnerabilities in each asset, it can be said that \texttt{libssl} is the most vulnerable asset in this version. Nevertheless, it is never enough to analyze the number of vulnerabilities ($M_3$), and the CVSS score for each has to be taken into account, as well as the existence of known exploits (\textit{e.g.}, updating the value of the CVSS using the temporal score).
                \item The other vulnerabilities in OpenPLC V1 affect the assets of the operating system, such as \texttt{libc} ($M_3 = 9$).
                \item The \textit{ad hoc} assets developed for this project have no known vulnerabilities. This does not mean that they are secure, but rather that there are no known vulnerabilities available for them. Zero-day vulnerabilities could be present in the SUT.
            \end{itemize}
            
            Values of $M_1 = 91$ suggests a high number of vulnerabilities in this project. This is expected given that OpenPLC V1 was the first version of OpenPLC. In the following versions, the value of $M_1$ is expected to decrease.
            
            The most striking fact when observing the EDG for OpenPLC V1 is the large number of vulnerabilities that are present in \textit{libssl}. From this perspective, if a pen-tester were to evaluate OpenPLC V1, \textit{libssl} would be a promising target. When the values of the CVSS are checked, three vulnerabilities\footnote{\href{https://www.cvedetails.com/cve/CVE-2016-2842/}{CVE-2016-2842},  \href{https://www.cvedetails.com/cve/CVE-2016-0799/}{CVE-2016-0799}, and  \href{https://www.cvedetails.com/cve/CVE-2016-0705/}{CVE-2016-0705}.} have a score of 10.0 out of 10.0. As can be seen, EDGs are a powerful tool for inspecting the structure of the SUT, and they can be used to analyze how the exploitation of a vulnerability can affect the rest of the SUT.
    
            All of the assets point to \textit{libc}, which is a wrapper around the system calls of the Linux kernel. An evaluator, even without this information, can understand the importance of this dependency with \textit{libc} using the EDG. On closer inspection, the highest score of CVSS for the vulnerabilities of \textit{libc} is 9.3 out of 10.0\footnote{\href{https://www.cvedetails.com/cve/CVE-2017-16997/}{CVE-2017-16997}}. If the exploitation of this vulnerability is possible, then it would allow a local user to gain privileges, exposing other assets of the SUT. 
            
            Using the value of severity of each vulnerability, it is possible to generate a list of vulnerabilities to be patched. This list can either be ordered by the global CVSS or by asset. Table \ref{tab:OpenPLCv1Prioritization} shows all the vulnerabilities whose CVSS value is between 6.0 and 10.0 ordered by asset and by descending CVSS. This can be used to decide in which order the vulnerabilities have to be patched, both at asset level or at SUT level.

            \begin{table}[!htb]
            \centering
            \caption{Vulnerability prioritization by asset and by CVSS for OpenPLC V1. Only the vulnerabilities whose value is between 6.0 and 10.0 are shown.}
            \label{tab:OpenPLCv1Prioritization}
            \resizebox{0.3\textwidth}{!}{%
            \begin{tabular}{lcl}
            \hline
            \multicolumn{1}{c}{CVE} & \multicolumn{1}{c}{CVSS} & \multicolumn{1}{l}{ASSET} \\ \hline
            CVE-2019-15847          & 7.5                      & libcares                  \\ \hline
            CVE-2018-12886          & 7.5                      & libgcc                    \\ \hline
            CVE-2016-9843           & 7.5                      & libz                      \\
            CVE-2016-9841           & 7.5                      & libz                      \\
            CVE-2016-9840           & 6.8                      & libz                      \\
            CVE-2016-9842           & 6.8                      & libz                      \\ \hline
            CVE-2017-16997          & 9.3                      & libc                      \\
            CVE-2014-9984           & 7.5                      & libc                      \\
            CVE-2014-4043           & 7.5                      & libc                      \\
            CVE-2015-5277           & 7.2                      & libc                      \\
            CVE-2015-7547           & 6.8                      & libc                      \\
            CVE-2014-0475           & 6.8                      & libc                      \\ \hline
            CVE-2016-2842           & 10.0                     & libssl                    \\
            CVE-2016-0705           & 10.0                     & libssl                    \\
            CVE-2016-0799           & 10.0                     & libssl                    \\
            CVE-2016-6304           & 7.8                      & libssl                    \\
            CVE-2016-0798           & 7.8                      & libssl                    \\
            CVE-2014-8176           & 7.5                      & libssl                    \\
            CVE-2016-2182           & 7.5                      & libssl                    \\
            CVE-2014-3512           & 7.5                      & libssl                    \\
            CVE-2016-6303           & 7.5                      & libssl                    \\
            CVE-2015-0292           & 7.5                      & libssl                    \\
            CVE-2016-2177           & 7.5                      & libssl                    \\
            CVE-2014-3567           & 7.1                      & libssl                    \\
            CVE-2014-3513           & 7.1                      & libssl                    \\
            CVE-2015-1791           & 6.8                      & libssl                    \\
            CVE-2012-2333           & 6.8                      & libssl                    \\
            CVE-2015-0209           & 6.8                      & libssl                    \\
            CVE-2014-3509           & 6.8                      & libssl                    \\
            CVE-2014-0195           & 6.8                      & libssl                    \\
            CVE-2014-3505           & 5.0                      & libssl                    \\ \hline
            \end{tabular}%
            }
            \end{table}
            
            From the information in Table \ref{tab:OpenPLCv1Prioritization}, it is clear that all the three vulnerabilities with a CVSS value of $10.0$ are in \texttt{libssl}, which makes this asset the most vulnerable by number of vulnerabilities and by CVSS. These vulnerabilities should be a priority during the patching stages. It is worth noting that \texttt{libc}, which is an important asset in Ubuntu Linux, has a vulnerability whose score is $9.3$ (CVE-2017-16997). This should also be a priority when patching.

            Moving now to OpenPLC V2, and comparing it with OpenPLC V1 (temporal dimension), similar conclusions can be drawn:

            \begin{itemize}
                \item Most of the vulnerabilities are in third-party open-source components. Moreover, \texttt{libssl} ($M_3 = 63$) remains as the asset with the majority of vulnerabilities (with $M_4 = 81.8 \%$ of them), followed by \texttt{libz} ($M_3 = 4$) from \texttt{node-js}.
                \item The remaining vulnerabilities are related to the operating system, just as in OpenPLC V1, \texttt{libc} ($M_3 = 5$).
            \end{itemize}

            For this version, $M_1 = 77$, less than $M_1 = 98$ for the previous version. As was expected, the total number of vulnerabilities has decreased. From the \textbf{spatial dimension}, it is worth noting that \texttt{libssl} is still the asset with the highest number of vulnerabilities and is still a promising target for a pen-tester or an attacker. Nevertheless, the number of vulnerabilities for \texttt{libssl} ($M_3$) has decreased from OpenPLC V1 to OpenPLC V2, as it was expected.
            
            As was done earlier, a list of vulnerabilities can be generated that is ordered by asset and by descending CVSS. Table \ref{tab:OpenPLCv2Prioritization} shows all of the vulnerabilities for OpenPLC V2 whose CVSS value is between 6.0 and 10.0.

            \begin{table}[!htb]
            \centering
            \caption{Vulnerability prioritization by asset and by CVSS for OpenPLC V2. Only the vulnerabilities whose value is between 6.0 and 10.0 are shown.}
            \label{tab:OpenPLCv2Prioritization}
            \resizebox{0.3\textwidth}{!}{%
            \begin{tabular}{lcl}
            \hline
            \multicolumn{1}{c}{CVE} & \multicolumn{1}{c}{CVSS} & \multicolumn{1}{l}{ASSET} \\ \hline
            CVE-2018-12886          & 6.8                      & libgcc\_s                 \\ \hline
            CVE-2017-16997          & 9.3                      & libc                      \\
            CVE-2017-18269          & 7.5                      & libc                      \\ \hline
            CVE-2016-9843           & 7.5                      & libz                      \\
            CVE-2016-9841           & 7.5                      & libz                      \\
            CVE-2016-9842           & 6.8                      & libz                      \\
            CVE-2016-9840           & 6.8                      & libz                      \\ \hline
            CVE-2016-0799           & 10.0                     & libssl                    \\
            CVE-2016-2108           & 10.0                     & libssl                    \\
            CVE-2016-0705           & 10.0                     & libssl                    \\
            CVE-2016-2842           & 10.0                     & libssl                    \\
            CVE-2016-0798           & 7.8                      & libssl                    \\
            CVE-2016-6304           & 7.8                      & libssl                    \\
            CVE-2016-2109           & 7.8                      & libssl                    \\
            CVE-2016-2177           & 7.5                      & libssl                    \\
            CVE-2016-2182           & 7.5                      & libssl                    \\
            CVE-2016-6303           & 7.5                      & libssl                    \\
            CVE-2015-0209           & 6.8                      & libssl                    \\
            CVE-2015-1791           & 6.8                      & libssl                    \\
            CVE-2016-2106           & 6.5                      & libssl                    \\
            CVE-2016-2176           & 6.4                      & libssl                    \\ \hline
            \end{tabular}%
            }
            \end{table}
            
            Table \ref{tab:OpenPLCv2Prioritization} shows that \texttt{libssl} now has four instead of three vulnerabilities whose CVSS value is 10.0 (CVE-2016-0799, CVE-2016-2108, CVE-2016-0705, CVE-2016-2842). In OpenPLC V2, \texttt{libc} is still affected by the same vulnerability with a score of 9.3 (CVE-2017-16997). All of these vulnerabilities should be a priority during patching activities.

            Finally, OpenPLC V3, which is the last available version, is analyzed. The most striking fact when comparing OpenPLC V3 with the other versions (\textbf{temporal dimension}) is the significant reduction in the global number of vulnerabilities. This effect is caused by the absence of \texttt{libssl} in this version.
            
            Focusing on the \textbf{spatial dimension} for this version, vulnerabilities in OpenPLC V3 are only related to assets in the operating system, \textit{libgcc\_s} ($M_3 = 2$), and \textit{libc} ($M_3 = 3$). In OpenPLC V3, neither third-party open-source assets nor \textit{ad hoc} assets have any known vulnerability. In fact, OpenPLC V3 has the lowest values of $M_1 = 5$ for the whole life cycle of the project, and therefore has the lowest number of vulnerabilities.
            
            Table \ref{tab:OpenPLCv3Prioritization} shows an ordered list of all the vulnerabilities in OpenPLC V3 whose CVSS value is between 6.0 and 10.0.

            \begin{table}[!htb]
            \centering
            \caption{Vulnerability prioritization by asset and by CVSS for OpenPLC V3. Only the vulnerabilities whose value is between 6.0 and 10.0 are shown.}
            \label{tab:OpenPLCv3Prioritization}
            \resizebox{0.3\textwidth}{!}{%
            \begin{tabular}{lcl}
            \hline
            \multicolumn{1}{c}{CVE} & \multicolumn{1}{c}{CVSS} & \multicolumn{1}{l}{ASSET} \\ \hline
            CVE-2018-12886          & 6.8                      & libgcc\_s                 \\ \hline
            CVE-2018-11236          & 7.5                      & libc                      \\
            CVE-2017-18269          & 7.5                      & libc                      \\ \hline
            \end{tabular}%
            }
            \end{table}
            
            For OpenPLC V3, the most vulnerable asset is \texttt{libc}, according to the CVSS score shown in Table \ref{tab:OpenPLCv3Prioritization}.

        \subsubsection{Root Causes Analysis (Weaknesses)}
            In this last step of the analysis, the information about the root causes of the vulnerabilities is extracted. This can be achieved through weaknesses. Each CWE contains data about the main issue to be solved for any given vulnerability. This information is useful to propose new requirements, test cases, and training. The analysis of the root causes will also be carried out in both the spatial and temporal dimensions.
            
\begin{figure}[!htb]
              \begin{center}
              \includegraphics[width=3.4in]{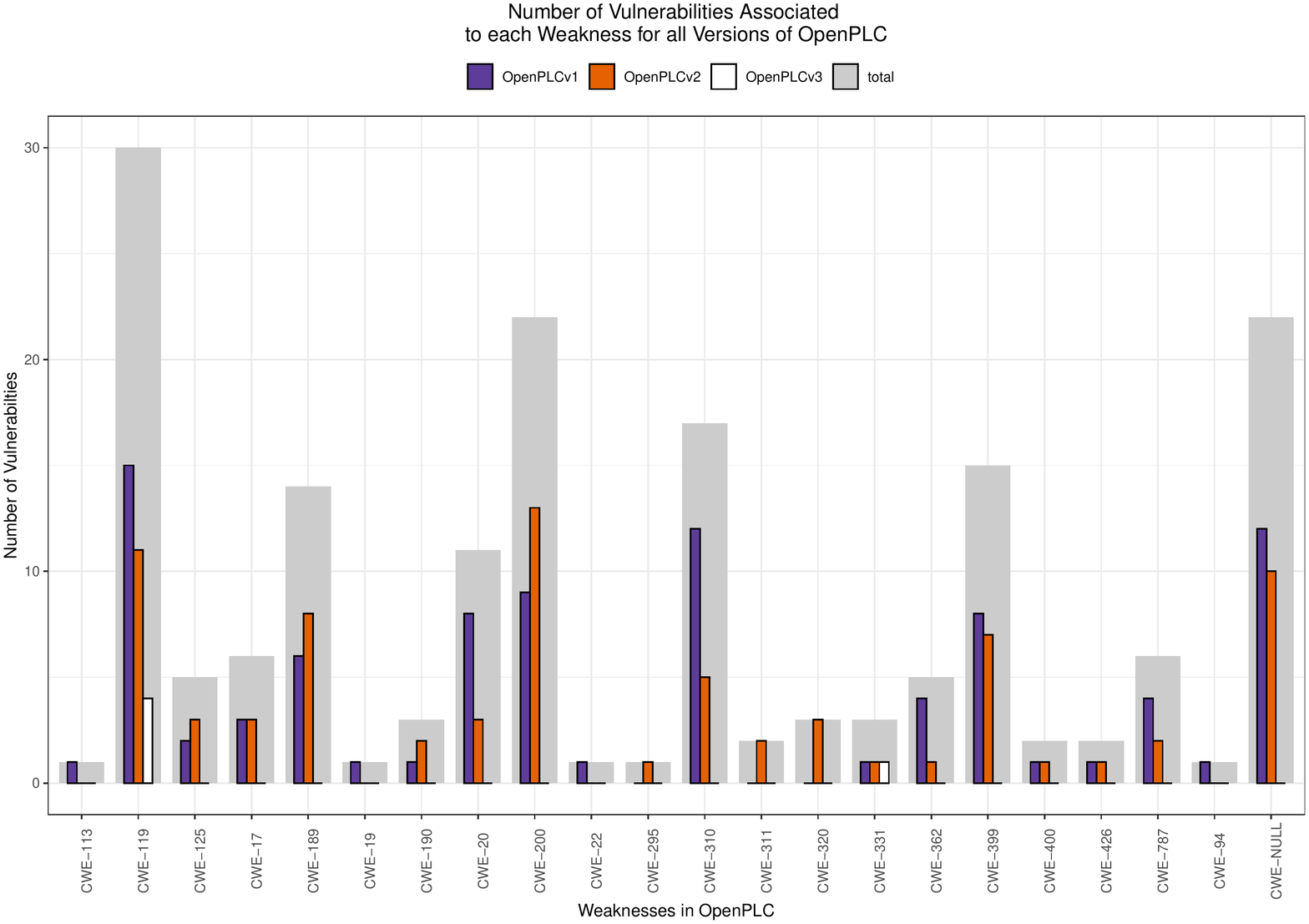}
              \caption{Evolution of the number of vulnerabilities over consecutive versions of OpenPLC.}
              \label{fig:openPLC_weaknesses}
              \end{center}
            \end{figure}

            The analysis of the root causes starts by extracting the weaknesses for each vulnerability for all three versions of OpenPLC. All of the extracted weaknesses for all three available versions of OpenPLC are shown in Fig. \ref{fig:openPLC_weaknesses} and in Table \ref{tab:SUTweaknesses}.

            % % =======
            % % FIG. XX
            % % =======
            % \begin{figure}[!htb]
            %   \begin{center}
            %   \includegraphics[width=3.4in]{photo/OpenPLC_v_t.pdf}
            %   \caption{Evolution of the number of vulnerabilities over consecutive versions of OpenPLC.}
            %   \label{fig:openPLC_weaknesses}
            %   \end{center}
            % \end{figure}
            % He movido esta figura más abajo para ajutar mejor el texto
            
\begin{table}[!htb]
\centering
\caption{Weakness analysis for all versions of OpenPLC.}
\label{tab:SUTweaknesses}
\resizebox{0.5\textwidth}{!}{%
\begin{tabular}{lrcllllllclllllcc}
\hline
\multicolumn{2}{c}{METRIC} & \multicolumn{7}{c}{OpenPLC V1} & \multicolumn{6}{c}{OpenPLC V2} & OpenPLC V3 & \multicolumn{1}{l}{$\sum_{i=1}^{T}M_8(A, cwe_j, t)$} \\ \hline
\multirow{22}{*}{\rotatebox{90}{$M_{8}( A, cwe_j, t ) = \vert CVE_{A\vert cwe_j}(t)\vert$}} & CWE-17 & \multicolumn{7}{c}{3} & \multicolumn{6}{c}{3} & - & 6 \\
 & CWE-19 & \multicolumn{7}{c}{1} & \multicolumn{6}{c}{-} & - & 1 \\
 & CWE-20 & \multicolumn{7}{c}{8} & \multicolumn{6}{c}{3} & - & 11 \\
 & CWE-22 & \multicolumn{7}{c}{1} & \multicolumn{6}{c}{-} & - & 1 \\
 & CWE-94 & \multicolumn{7}{c}{1} & \multicolumn{6}{c}{-} & - & 1 \\
 & CWE-113 & \multicolumn{7}{c}{1} & \multicolumn{6}{c}{-} & - & 1 \\
 & CWE-119 & \multicolumn{7}{c}{15} & \multicolumn{6}{c}{11} & 4 & 30 \\
 & CWE-125 & \multicolumn{7}{c}{2} & \multicolumn{6}{c}{3} & - & 5 \\
 & CWE-189 & \multicolumn{7}{c}{6} & \multicolumn{6}{c}{8} & - & 14 \\
 & CWE-190 & \multicolumn{7}{c}{1} & \multicolumn{6}{c}{2} & - & 3 \\
 & CWE-200 & \multicolumn{7}{c}{9} & \multicolumn{6}{c}{13} & - & 22 \\
 & CWE-295 & \multicolumn{7}{c}{-} & \multicolumn{6}{c}{1} & - & 1 \\
 & CWE-310 & \multicolumn{7}{c}{12} & \multicolumn{6}{c}{5} & - & 17 \\
 & CWE-311 & \multicolumn{7}{c}{-} & \multicolumn{6}{c}{2} & - & 2 \\
 & CWE-320 & \multicolumn{7}{c}{-} & \multicolumn{6}{c}{3} & - & 3 \\
 & CWE-331 & \multicolumn{7}{c}{1} & \multicolumn{6}{c}{1} & 1 & 3 \\
 & CWE-362 & \multicolumn{7}{c}{4} & \multicolumn{6}{c}{1} & - & 5 \\
 & CWE-399 & \multicolumn{7}{c}{8} & \multicolumn{6}{c}{7} & - & 15 \\
 & CWE-400 & \multicolumn{7}{c}{1} & \multicolumn{6}{c}{1} & - & 2 \\
 & CWE-426 & \multicolumn{7}{c}{1} & \multicolumn{6}{c}{1} & - & 2 \\
 & CWE-787 & \multicolumn{7}{c}{4} & \multicolumn{6}{c}{2} & - & 6 \\
 & CWE-NULL & \multicolumn{7}{c}{12} & \multicolumn{6}{c}{10} & - & 22 \\ \hline
\multicolumn{2}{l}{$M_7( A, t ) = \vert CWE_A(t) \vert$} & \multicolumn{7}{c}{19} & \multicolumn{6}{c}{18} & 2 & 22 \\ \hline
\end{tabular}%
}
\end{table}
            
            Focusing on the spatial dimension, OpenPLC V1 has the widest range of weaknesses ($M_7 = 19$), and therefore has the widest range of root causes. From the total number of vulnerabilities in OpenPLC V1 ($M_3 = 91$), $M_6 = 15$ of them are related to weaknesses CWE-119\footnote{Description for CWE-119: \url{https://cwe.mitre.org/data/definitions/119.html}}: ``The software performs operations on a memory buffer, but it can read from or write to a memory location that is outside the intended boundary of the buffer.'' $M_6 = 12$ of them are related to CWE-310 \footnote{Description for CWE-310: \url{https://cwe.mitre.org/data/definitions/310.html}.}: ``Weaknesses in this category are related to the design and implementation of data confidentiality and integrity. Frequently, these deal with the use of encoding techniques, encryption libraries, and hashing algorithms. The weaknesses in this category could lead to a degradation of the quality of data if they are not addressed.''

            Repeating this same analysis with OpenPLC V2, we can see that the number of different weaknesses ($M_7 = 18$) is almost the same as before. But in this version, the most repeated weakness is CWE-200\footnote{Description for CWE-200: \url{https://cwe.mitre.org/data/definitions/200.html}}: ``The product exposes sensitive information to an actor that is not explicitly authorized to have access to that information.'' Followed by CWE-119.
            
            Comparing the previous versions with OpenPLC V3, we find now that the total number of weaknesses ($M_7 = 2$) has been reduced drastically. Nevertheless, weakness CWE-119 is still present in this version, and is the most repeated.

            This analysis can also be done by combining the information of all three versions, drawing conclusions for the whole life cycle of OpenPLC. Requirements and test cases can be extracted from the most recurrent weaknesses, and training can be proposed to avoid weaknesses that repeat over time. In our example, CWE-119 has the greatest frequency over time ($\sum_{i=1}^{T}M_6 = 30$), followed by CWE-200 ($\sum_{i=1}^{T}M_6 = 22$), and CWE-310 ($\sum_{i=1}^{T}M_6 = 17$). Table \ref{tab:requirements} shows the generated requirements, Table \ref{tab:training} shows the proposed training, and Table \ref{tab:testCases} shows the proposed test cases for OpenPLC.

            % =======
            % FIG. XX
            % =======

\begin{table*}[!htb]
\centering
\caption{An example of generated requirements for OpenPLC.}
\label{tab:requirements}
\resizebox{\textwidth}{!}{%
\begin{tabular}{@{}p{0.25\linewidth}p{0.75\linewidth}@{}}
\toprule
CWE ID &
  REQUIREMENTS \\ \midrule
CWE-19, CWE-20, CWE-119, CWE-125, CWE-189, CWE-190, CWE-362, CWE-399, CWE-400, CWE-787 &
  Use languages that perform their own memory management. \\
 &
  \\
 CWE-19, CWE-20, CWE-119, CWE-125, CWE-189, CWE-190, CWE-787 &
  Use libraries or frameworks that make it easier to handle numbers without unexpected consequences. Examples include safe integer handling packages such as SafeInt (C++) or IntegerLib (C or C++). \\
 &
  \\
 CWE-19, CWE-20, CWE-119, CWE-125, CWE-189, CWE-190, CWE-200, CWE-362, CWE-787 &
  Use a CPU and operating system that offers Data Execution Protection (NX) or its equivalent. \\
 &
  \\
 CWE-19, CWE-20, CWE-125, CWE-189, CWE-190, CWE-200, CWE-362 &
  Ensure that all protocols are strictly defined, such that all out-of-bounds behaviors can be identified simply, and require strict conformance to the protocol. \\
 &
  \\
 CWE-19, CWE-20, CWE-399, CWE-400 &
  Run or compile the software using features or extensions that randomly arrange the positions of a program's executable and libraries in memory (\textit{e.g.}, Address Space Layout Randomization (ASLR), or Position-Independent Executables (PIE)). \\
 &
  \\
 CWE-19, CWE-295, CWE-310, CWE-311, CWE-320, CWE-331 &
  Clearly specify which data or resources are valuable enough that they should be protected by encryption. Require that any transmission or storage of this data/resource should use well-vetted encryption algorithms. Up-to-date algorithms must be used, and the entropy of the keys must be sufficient for the application. \\
 &
  \\
 CWE-22, CWE-94, CWE-113, CWE-399, CWE-400, CWE-426 &
  Hard-code the search path to a set of known-safe values (such as system directories), or only allow them to be specified by the administrator in a configuration file. Do not allow these settings to be modified by an external party. \\
 &
  \\
 CWE-20, CWE-22, CWE-94, CWE-113, CWE-399, CWE-400, CWE-426 &
  Use an input validation framework such as Struts or the OWASP ESAPI Validation API. \\
 &
  \\
 CWE-19, CWE-20, CWE-22, CWE-94, CWE-113, CWE-295, CWE-426 &
  Assume all input is malicious. Use an "accept known good" input validation strategy, i.e., use a list of acceptable inputs that strictly conform to specifications. Reject any input that does not strictly conform to specifications, or transform it into something that does. \\
 &
  \\
 CWE-19, CWE-20, CWE-119, CWE-125, CWE-399, CWE-400, CWE-787 &
  Run or compile the software using features or extensions that automatically provide a protection mechanism that mitigates or eliminates buffer overflows. \\
 &
  \\
 CWE-19, CWE-22, CWE-119, CWE-125, CWE-399, CWE-400, CWE-787 &
  Replace unbounded copy functions with analogous functions that support length arguments, such as strcpy with strncpy. Create these if they are not available. \\ \bottomrule
\end{tabular}%
}
\end{table*}

\begin{table*}[!htb]
\centering
\caption{Example of proposed training for OpenPLC.}
\label{tab:training}
\resizebox{\textwidth}{!}{%
\begin{tabular}{@{}p{0.25\linewidth}p{0.75\linewidth}@{}}
\toprule
CWE ID & TRAINING                                                                               \\ \midrule
CWE-19, CWE-20, CWE-22, CWE-94, CWE-113, CWE-119, CWE-125, CWE-426 &
  Identification of all potentially relevant properties of an input (length, type of input, the full range of acceptable values, missing or extra inputs, syntax, consistency across related fields). \\
&                                                                                        \\
CWE-19, CWE-20, CWE-22, CWE-94, CWE-113, CWE-119, CWE-125, CWE-426 & Input validation strategies.                                                            \\
      &                                                                                        \\
CWE-19, CWE-20, CWE-22, CWE-94, CWE-113, CWE-119, CWE-125, CWE-200, CWE-426 & Allowlists and Denylists.                                                               \\
      &                                                                                        \\
CWE-19, CWE-20, CWE-22, CWE-94, CWE-113, CWE-119, CWE-125, CWE-189, CWE-190, CWE-426 & Character encoding compatibility.                                                       \\
      &                                                                                        \\
CWE-19, CWE-20, CWE-94, CWE-113, CWE-119, CWE-125, CWE-426 &
  Buffer overflow detection during compilation (\textit{e.g.}, Microsoft Visual Studio /GS flag, Fedora/Red Hat FORTIFY\_SOURCE GCC flag, StackGuard, and ProPolice). \\
      &                                                                                        \\
CWE-19, CWE-20, CWE-94, CWE-113, CWE-119, CWE-125, CWE-200, CWE-426 & Secure functions, such as \textit{strcpy} with \textit{strncpy}. Create these if they are not available. \\
      &                                                                                        \\
CWE-19, CWE-20, CWE-94, CWE-113, CWE-119, CWE-125, CWE-189, CWE-190, CWE-426,CWE-787 & Secure programming: memory management.                                                 \\
      &                                                                                        \\
CWE-19, CWE-20, CWE-22, CWE-94, CWE-113, CWE-119, CWE-125, CWE-189, CWE-190, CWE-787 &
  Understand the programming language's underlying representation and how it interacts with numeric calculation. \\
      &                                                                                        \\
CWE-19, CWE-20, CWE-22, CWE-94, CWE-113, CWE-119, CWE-125 & System compartmentalization.                                                            \\
      &                                                                                        \\
CWE-200, CWE-295, CWE-310, CWE-311, CWE-320, CWE-331 & Certificate management.                                                                 \\
      &                                                                                        \\
CWE-200, CWE-295, CWE-310, CWE-311, CWE-320, CWE-331 & Certificate pinning.                                                                    \\
      &                                                                                        \\
CWE-310, CWE-311, CWE-320, CWE-331 & Encryption integration (Do not develop custom or private cryptographic algorithms).     \\
      &                                                                                        \\
CWE-310, CWE-311, CWE-320, CWE-331 & Secure up-to-date cryptographic algorithms.                                             \\
      &                                                                                        \\
CWE-189, CWE-200, CWE-362, CWE-399, CWE-400, CWE-426 & Shared resources management.                                                            \\
      &                                                                                        \\
CWE-200, CWE-362, CWE-399, CWE-400, CWE-426 & Thread-safe functions.                                                                        \\
      &                                                                                        \\ \bottomrule
\end{tabular}%
}
\end{table*}

\begin{table*}[!htb]
\centering
\caption{Example of generated test cases for OpenPLC.}
\label{tab:testCases}
\resizebox{\textwidth}{!}{%
\begin{tabular}{@{}p{0.25\linewidth}p{0.75\linewidth}@{}}
\toprule
CAPEC ID & TEST CASES                                                                     \\ \midrule
CAPEC-10   & Check for buffer overflows through manipulation of environment variables. This test leverages implicit trust often placed in environment variables.             \\
      &                                                                                 \\
CAPEC-14   & Static analysis of the code: secure functions and buffer overflow.             \\
      &                                                                                 \\
CAPEC-24   & Feed overly long input strings to the program in an attempt to overwhelm the filter (by causing a buffer overflow) and hoping that the filter does not fail securely (i.e. the user input is let into the system unfiltered)             \\
      &                                                                                 \\
CAPEC-45   & This test uses symbolic links to cause buffer overflows. The evaluator can try to create or manipulate a symbolic link file such that its contents result in out of bounds data. When the target software processes the symbolic link file, it could potentially overflow internal buffers with insufficient bounds checking.             \\
      &                                                                                 \\
CAPEC-46   & Static analysis of the code: secure functions and buffer overflow.             \\
      &                                                                                 \\
CAPEC-47   & In this test, the target software is given input that the evaluator knows will be modified and expanded in size during processing. This test relies on the target software failing to anticipate that the expanded data may exceed some internal limit, thereby creating a buffer overflow.             \\
      &                                                                                 \\
CAPEC-59   & This test targets predictable session ID in order to gain privileges. The evaluator can try to predict the session ID used during a transaction to perform spoofing and session hijacking.             \\
      &                                                                                 \\
CAPEC-97   & Automated measurement of the entropy of the keys generated.             \\
      &                                                                                 \\
CAPEC-475   & Fuzz testing to externally provided data, such as directories and filenames.   \\
      &                                                                                 \\ \bottomrule
\end{tabular}%
}
\end{table*}

\section{Conclusions and Future Work}
\label{sec:conclusionsFutureWork}
    Industrial components have become the driving force behind the development of modern industry. They are a key element in many sectors. They perform critical tasks and they process critical data. Following the introduction of connectivity, the use of COTS and open-source components, and the exponentially increasing complexity of industrial components, the number of threats to which industrial components are exposed has increased. These factors, together with the large lifespan of industrial components, call for a vulnerability assessment methodology that monitors the entire life cycle of industrial components. Existing security standards agree that vulnerability assessment is a critical task, but they usuallyonly consider software in their analysis, when both software and hardware should be taken into account.

    The EDG model that is proposed in this research work constitutes a method to assess known vulnerabilities that is aligned with the ISA/IEC 62443 standard. EDGs have been specially developed to assess vulnerabilities in industrial components, but they can also be applied to any SUT that can be decomposed into assets. The key feature of the proposed model is the addition of the temporal dimension in the analysis of vulnerabilities. This makes it possible to analyze the location of vulnerabilities in both space (in which asset) and time (their recurrence), which allows the state of the device to be tracked throughout the whole life cycle. In addition, metrics constitute a plethora of information that that can be used by the model to improve the development process of the SUT, enhance its security, and track its status during its whole life cycle. The proposed EDG model has successfully achieved the integration of the following features:

    \begin{itemize}
        \item Directed graph-based analysis model.
        \item Temporal evolution of the SUT, its assets, and its vulnerabilities.
        \item Quantitative metrics to measure the state of the SUT over time.
        \item Assess vulnerabilities in both software and hardware.
    \end{itemize}
    
    Moreover, with the information gathered by the model, it is possible to assist in update management activities, apply patching policies, launch training activities, and generate new test cases and requirements. This model can serve as a starting point for pen-testers, because it identifies vulnerabilities, weaknesses, and attack patterns for a given SUT.

    Finally, the OpenPLC use case served as a demonstration of the advantages of the EDG model, its applicability, and the potential of the model. The generated EDG for all three OpenPLC versions serve as a visual tool for the evaluators, helping them to know the preliminary status of the SUT, before any further analysis. Furthermore, the proposed model is able to explore internal dependencies to understand how a vulnerability would propagate throughout the SUT. We also found that the main root causes for the vulnerabilities in OpenPLC are related to buffer management operations, and cryptographic issues. With this data, we are able to propose new requirements, test cases, and training. Lists of vulnerability prioritizations for each version were also generated.

    As a future work, the EDG model will be enhanced by adding a mathematical model to aggregate the values of the CVSS metric for each asset, and a value for the whole SUT. This would allow us to compare different SUTs over time. More improvements will be done in the prioritization of patching, taking into account the context and the functionalities of the SUT. Finally, historical information about the developers can be integrated into the EDG model to predict future vulnerabilities. 
    % Finally, this approach is prone to automation, reducing the effort necessary to generate the model and update the data over time. Alerts and warnings can also be programmed in the case of excessively high CVSS values, or metrics outside established bounds.
% \input{docs/8.Appendixes}

% =================================================================================
% === BIBLIOGRAPHY ================================================================
% =================================================================================
\bibliographystyle{ieeetr}
\bibliography{references}

\end{document}